\documentclass[11pt,letterpaper]{article}
\usepackage[top=0.85in,left=0.75in,footskip=0.75in]{geometry}
\usepackage[utf8]{inputenc}
\usepackage{authblk}

\usepackage{natbib}
\usepackage{graphicx}
\usepackage{float}
\usepackage{amsmath}
\usepackage{amssymb}
\usepackage{xcolor}
\usepackage{natbib}

\title{Griffiths phase in clique percolation in random geometric graphs}
\author[1,2,*]{Vasilii Tiselko}
\author[1,3]{Olga Valba}
\author[1,2,4]{Alexander Gorsky}

\affil[1]{Laboratory of Complex Networks, Center for Neurophysics and Neuromorphic Technologies, Moscow, Russia}
\affil[2]{Moscow Institute for Physics and Technology, Phystech School of Applied Mathematics and Computer Science, Dolgoprudny 141700, Russia}
\affil[3]{Department of Applied Mathematics, MIEM, National Research University Higher School of Economics, Moscow, Russia}
\affil[4]{Kharkevich Institute for Information Transmission Problems RAS,  Moscow, Russia}
\affil[*] {vasily.tiselko@gmail.com}
\date{}                     
\makeatletter
\renewcommand\@biblabel[1]{#1.}
\makeatother

\usepackage{hyperref}
\hypersetup{colorlinks=true,linkcolor=blue,urlcolor=magenta,citecolor=blue}

\begin{document}
\maketitle

%
%\tableofcontents

\begin{abstract}
In this study, we discuss the clique percolation in the ensembles of random geometric graphs with different kernels that quantify the geometrical constraints. For the sharp cut-off we find the wide Griffiths phase of extended criticality with the power-law behavior. One boundary of the Griffiths phase is the generalization of a percolation critical point for the ER ensemble when the percolation within the large but finite cluster emerges. The second boundary corresponds to the point in the parameter space when the percolation in the entire clustered system becomes available. For the power-law kernel, richer behavior with a clique-size-dependent boundary between the effective ER and geometric regimes has been identified. The Griffiths phase in this case exists as well. Finally, the pattern with the exponential kernel has been analyzed. We briefly discuss the possible applications of our findings.

\end{abstract}

\newpage
\tableofcontents
\newpage

\section{Introduction}

Clique percolation is a clear and general phenomenon that generalizes the standard percolation
of bonds, which is one of the simplest critical phenomena \cite{derenyi2005clique,palla2007critical,bollobas2009clique,li2015clique, li2021percolation, goltsev2006k}. Its applications range from transport networks and community detection to general cognitive  networks \cite{stella2022cognitive,valba2022k}, 
or large-scale human connectomes \cite{tiselko2022k}. Certainly, the details
of the percolation of clusters depend on the structure of the underlying network manifold. It could be
the kind of network, graphs of system of interacting constituents.

There are critical parameters of the systems that make percolation possible. 
The familiar bond-percolation threshold in the Erd\H{o}s--R\'enyi network
ensemble is the edge probability \(p_c\simeq N^{-1}\), equivalently the
critical mean degree \(q_c\simeq1\). 
The exact results for the percolation thresholds for the $k$ -cliques are available, generalizing
the result for the bond percolation \cite{derenyi2005clique}. However, in a general ensemble of graphs, the analytical evaluation
of the critical parameter for the percolation threshold is not available, and 
numerical tools are applied. 

 In this study, we consider the percolation of k-clique in random geometric graphs(RGG)  when the probability of the link between
 nodes depends on the distance between them \cite{gilbert1961random,penrose2003random,dall2002random,duchemin2023random}. Such framework allows to account effectively for the range of interaction between the constituents of underlying manifold or media.
The short-range of interaction  is introduced by
the sharp cut-off or exponential behavior $\exp(-mr)$ when the intrinsic geometrical scale $m^{-1}$ provides the cut-off.
The large-range interaction is quantified by the power-law kernel. The dependence on the cut-off kernel 
has been investigated in \cite{duchemin2023random}. Some specific criticality in random geometric graphs has been found in \cite{ostilli2015statistical},
while aspects of clustering were investigated in \cite{sonmez2024distances, havlin1984topological}. Geometric restrictions
are used for the effective description and generative models of human connectomes \cite{gabay2017cortical,roberts2017consistency,pang2023geometric,bobyleva2025metric}.

We have found that remarkably the k-clique percolation in RGG manifests the Griffiths phase of extended criticality with the power-law behavior of observables.  the boundaries of the Griffiths phase have a clear-cut description. One boundary is the analog of the critical point of percolation in the ER ensemble. 
%%%
However, in the RGG case it describes percolation in local finite clusters. At the second boundary identified by the divergence of the particular susceptibility, the k-clique percolation becomes available throughout the network. In the Erd\H{o}s--R\'enyi limit, these two criticalities occur simultaneously: local self-sustained clique continuation already implies the emergence of system-scale clique percolation. Note that Griffiths phases have been observed in other critical phenomena on the networks. It was argued in \cite{munoz2010griffiths} that the Griffiths phase in networks is a fairly general phenomenon when there are long-lived clusters in networks with some kind of disorder.
In \cite{moretti2013griffiths, odor2019critical, pretel2024asynchronous} the Griffiths phases were observed in the synchronization phase transitions in the artificial and real networks. 
On the other hand, the extended single particle extended Anderson localization transition has been found in the full graph with diagonal disorder \cite{kravtsov2015random}, hierarchical modular networks \cite{odor2015griffiths} connectomes \cite{odor2022differences, bobyleva2025metric}, or even in
a deterministic system with time reversal symmetry breaking \cite{gorsky2026theta}. The review of the Griffiths phase can be found in \cite{vojta2006rare}.

We have found analytically  the local and global criteria which determine the boundaries of the Griffiths phase.
We also discuss that the critical mean degree depends non-monotonically on the dimension.  Its minimum occurs in intermediate dimensions, around \(d=4\)..\(6\), which is explained by how critical finite components merge with each other. However, in spite of the generic pattern of extended criticality, there is an essential dependence on the cut-off kernel. For the short-range interaction with sharp cut-off, we have clear-cut separation between the ER and the regime determined by the geometric constraint. In the power-law kernel the situation is more subtle, and the boundaries between regimes depend strongly on the details of the kernel and the size of the clique.
For the power-law kernel, we find an additional clique-size-dependent hierarchy of marginal boundaries between regimes, separating effective ER-like behavior from geometric locality.  Geometric correlations and locality appear first for larger clique events and only later for edges, so the crossover becomes clique-size dependent.  This gives an ordered ladder of marginal regimes with scale-free behavior in characteristic lengths emerging at the corresponding crossover scales.

The paper is organized as follows. In Section 2 we analytically formulate the general local and global criteria for the k-clique percolation. In Section 3 we discuss analytically and numerically the extended criticality for the sharp cutoff kernel. Section 4 is devoted to the discussion of the power-law kernel, while the exponential kernel is analyzed in Section 5.
The results of the paper and the open questions are summarized in the Discussion.

\section{General derivation of the local $k$-clique continuation criterion}

\subsection{Universal branching criterion and local criticality}

We consider a random geometric graph $G=G(q)$ on $N$ vertices with positions
$x_1,\ldots,x_N$ in a $d$-dimensional domain of volume $V$.  % and density $\rho=N/V$.%
Depending on positions, two vertices at distance $r$ are
connected with probability $W(r)$, where $W:[0,\infty)\to[0,1]$ is a radial
connection kernel. The random geometric graph with a sharp threshold corresponds to
$W(r)=\mathbf{1}_{r\le R}$, while soft kernels replace this sharp indicator by
a distance-dependent connection probability.

The goal is to describe the critical phenomena of clique percolation in the
presence of geometry, from the local birth of self-sustaining components to
percolation on the scale of the whole system.
The local criterion is derived by exploring the $k$-clique adjacency graph near
the onset of sparse connectivity.  At this scale, the exploration can be read as
a branching skeleton: a reached $k$-clique exposes forward $(k-1)$-faces, and
each such face may be completed by new vertices.  The tree-like picture is not
assumed to describe the full geometric transition; it is used only to identify
the point at which local clique continuation becomes self-sustaining at its
local onset.  The finite-component dynamics of macroscopic component formation
that follows this local onset is treated separately below.

The first object is therefore local criticality, not yet global percolation.
It can be understood as the stability of a branching exploration in the
$k$-clique adjacency graph, now constrained by geometry.  As in the
Galton--Watson picture, the local onset is reached when the average ability of
a discovered object to reproduce the next layer of the exploration becomes
one.

Let $\mathcal K_k(G)$ be the set of all $k$-cliques in $G$.  We form the
$k$-clique adjacency graph by declaring two $k$-cliques adjacent when they
share a common $(k-1)$-face.  The connected components of this adjacency graph
are the $k$-clique components.  For a $k$-clique component $\mathcal C$, its
clique mass is denoted by $M_{\mathcal C}$; this is the number of $k$-cliques
in $\mathcal C$, not the number of vertices.  This distinction is important
because a small set of vertices can carry many overlapping $k$-cliques.

Fix a $(k-1)$-face $Y=\{y_1,\ldots,y_{k-1}\}$.  A vertex $x\notin Y$ completes
this face in a $k$-clique if it connects to every vertex of $Y$.  The
completion probability is
\begin{equation}
\Pi_Y(x)=\prod_{y\in Y} W(\|x-y\|).
\label{eq:completion_probability}
\end{equation}
Let $C_Y$ be the number of vertices outside $Y$ that complete the face.  In
the standard Poisson spatial approximation, the conditional mean completion
count is
\begin{equation}
\lambda_Y=\rho\int \Pi_Y(x)\,\mathrm{d}x.
\label{eq:LambdaY}
\end{equation}
Here, $\rho=N/V$ is the vertex density. The finite-$N$ binomial version uses the same local mechanism with a finite
number of available outside vertices.  Equation~\eqref{eq:LambdaY} is the
large-system form that makes the geometry of the completion region explicit:
it is the expected size of the reservoir of vertices that can complete the
particular face \(Y\), with the kernel weighting each possible placement.

A local exploration of $k$-clique components does not see a uniformly chosen
$(k-1)$-face; it reaches a face through an already existing $k$-clique.  Faces
with many completions are therefore sampled more often because they
participate in more $k$-cliques.  The correct local continuation object is the
size-biased excess completion intensity
\begin{equation}
\Lambda_k=
\frac{\mathbb E[C_Y(C_Y-1)]}{\mathbb E[C_Y]}.
\label{eq:Lambda_k}
\end{equation}
Here, the expectation is taken over the distribution of reached
\((k-1)\)-faces in the local exploration, equivalently over an ensemble
of existing faces with the usual incidence bias generated by following an
already existing \(k\)-clique.
The subtraction by one has the same role as in the excess-degree calculation:
after a face has been reached through one completion, only the other
completions continue the exploration.

The same ratio has a direct finite-graph interpretation.  If
$N_k=|\mathcal K_k(G)|$ and the sum is over all $(k-1)$-cliques $Y$, then
the identity \(\sum_Y C_Y=kN_k\) shows that the denominator counts incidences
between $k$-cliques and their $(k-1)$-faces.  The corresponding numerator
\(\sum_Y C_Y(C_Y-1)\) counts ordered pairs of distinct $k$-cliques that share
the same $(k-1)$-face. Thus, $\Lambda_k$ is not a fitted parameter: it is the
expected number of alternative clique continuations seen from a face that has
already been reached through one existing $k$-clique.

If $C_Y$ is conditionally Poisson with mean $\lambda_Y$, then
$\mathbb E[C_Y(C_Y-1)\mid Y]=\lambda_Y^2$, and Eq.~\eqref{eq:Lambda_k} becomes
\begin{equation}
\Lambda_k=\frac{\mathbb E[\lambda_Y^2]}{\mathbb E[\lambda_Y]}.
\end{equation}
This is the $k$-clique analog of the excess degree in ordinary branching
percolation. Measures how many new clique continuations are typically
available from a face reached by following an existing clique.

A newly reached $k$-clique has $k$ different $(k-1)$-faces.  One face leads
back to the parent clique in local exploration; the other $k-1$ faces can
generate further continuations.  This gives the local continuation factor
\begin{equation}
B_k=(k-1)\Lambda_k.
\label{eq:branching}
\end{equation}
In the hard-threshold figures below, the running coordinate $q$ is the mean
degree of the underlying graph. In general formulas, $q$ denotes the
corresponding monotone scan coordinate of the random graph process, together
with the associated edge density of the network.
Equivalently, the early local exploration is the Galton--Watson
branching-process skeleton of the $k$-clique adjacency graph.  In this skeleton
, $B_k$ is the offspring number: each discovered clique has $k-1$ forward
faces, and each forward face contributes the mean excess completion intensity
$\Lambda_k$. The Galton--Watson criticality condition is the mean offspring one.
The local continuation onset is therefore defined by
\begin{equation}
q_B:\qquad B_k(q_B)=1.
\label{eq:percolation_criterion}
\end{equation}

The logic of this criterion is close to the classical branching derivation of
the $k$-clique percolation threshold in Erd\H{o}s--R\'enyi graphs
\cite{palla2007critical}.  The present formulation extends the same excess
continuation idea to geometric graphs and general kernels, where the completion
count depends on the geometry of the shared face.  In a fully mixed or
locally tree-like graph, this local onset is already the onset of macroscopic
$k$-clique connectivity.  In a random geometric graph, it is a local
criticality criterion and does not need to coincide with the criticality of forming a
percolating $k$-clique cluster.  Local clique continuation may become
self-sustaining while still being confined inside localized finite patches.
The finite-component coalescence layer responsible for the observable
transition is introduced below.

%%%%%%%%%%%%%%%%%%%%%% ER

\subsection{Erd\H{o}s--R\'enyi limit}

The Erd\H{o}s--R\'enyi graph is the reference case in which spatial
correlations are absent.  We denote its edge probability by $p^{\rm ER}$.  It
is the distance-independent kernel $W(r)\equiv p^{\rm ER}$.  For any fixed
$(k-1)$-face $Y$, a candidate vertex $x\notin Y$ must connect to all vertices
of $Y$, hence
\begin{equation}
\Pi_Y(x)=(p^{\rm ER})^{k-1}.
\end{equation}

There are $N-k+1$ vertices outside $Y$, so
$C_Y\sim{\rm Binomial}(N-k+1,(p^{\rm ER})^{k-1})$. Therefore
, $\mathbb E[C_Y]=(N-k+1)(p^{\rm ER})^{k-1}$ and
$\mathbb E[C_Y(C_Y-1)]=(N-k+1)(N-k)(p^{\rm ER})^{2(k-1)}$.  The excess
completion intensity is
\begin{equation}
\Lambda_k^{\rm ER}
=
\frac{\mathbb E[C_Y(C_Y-1)]}{\mathbb E[C_Y]}
=
(N-k)(p^{\rm ER})^{k-1}.
\end{equation}
Consequently,
\begin{equation}
B_k^{\rm ER}
=
(k-1)(N-k)(p^{\rm ER})^{k-1},
\end{equation}
and the finite-$N$ local-continuation condition is
\begin{equation}
(k-1)(N-k)(p_c^{\rm ER})^{k-1}=1.
\label{eq:ER_criterion}
\end{equation}
For fixed $k$ and large $N$, this gives the classical Erd\H{o}s--R\'enyi
$k$-clique percolation threshold
\begin{equation}
p_c^{\rm ER}(N,k)=\big[(k-1)N\big]^{-1/(k-1)}[1+O(k/N)].
\end{equation}

For $k=2$, this reduces to $p_c^{\rm ER}(N,2)\sim 1/N$, the standard
giant-component threshold.  For $k\ge 3$, the exponent $1/(k-1)$ reflects the
simultaneous completion constraint: a new vertex must connect to all vertices
of a prescribed $(k-1)$-face.  Thus, in the Erd\H{o}s--R\'enyi limit, local
criticality and global $k$-clique connectivity coincide asymptotically.  This
is precisely why Erd\H{o}s--R\'enyi is the natural reference point.  It is the
case where local branching already implies the emergence of a giant
percolating $k$-clique component, and no additional geometric finite-sector
layer is needed.

\subsection{Finite-component layer and geometric clique percolation criticality}

The size-biased excess intensity $\Lambda_k$ is the correct local branching
object at the level of a face reached.  Multiplying it by the $k-1$ forward
faces of a newly reached $k$-clique gives the mean offspring number
$B_k=(k-1)\Lambda_k$.  The criticality of this local branching does not need to
coincide with the criticality of forming a percolating $k$-clique cluster, and
the reason is precisely spatial localization, the mechanism absent in the
Erd\H{o}s--R\'enyi reference.

In the presence of geometric correlations, once local continuation becomes
possible, new $k$-cliques do not sample the whole graph independently.  They
remain near the region selected by the interaction kernel and tend to form
 finite localized patches.  These patches can grow internally and merge with
nearby patches before any component occupies a macroscopic fraction of the
system.

The random-geometric process therefore has two layers of criticality.  The
first is local continuation, detected by $B_k(q_B)=1$.  The second is the
finite-component layer, where localized clique components grow, merge, and are
eventually absorbed into the largest component.

Let $\Gamma_k(q)$ be the set of $k$-clique components of $G(q)$.  For
$\mathcal C\in\Gamma_k(q)$, let $M_{\mathcal C}(q)$ be its clique mass, i.e.
the number of $k$-cliques in the component.  The full second moment of
the component clique masses is
\begin{equation}
\chi_{\rm all}(q)
=
\sum_{\mathcal C\in\Gamma_k(q)}
M_{\mathcal C}(q)^2 .
\label{eq:chi_all_general}
\end{equation}
The largest clique-component mass is
\begin{equation}
M_{\max}(q)
=
\max_{\mathcal C\in\Gamma_k(q)}
M_{\mathcal C}(q),
\label{eq:mmax_general}
\end{equation}
and the finite-component susceptibility is obtained by removing the largest
component from the second moment:
\begin{equation}
\chi_{\rm finite}(q)
=
\chi_{\rm all}(q)-M_{\max}(q)^2 .
\label{eq:chi_finite_general}
\end{equation}

This is the clique-component analog of finite-cluster susceptibility in
ordinary percolation. Measures the second-moment mass stored outside the
dominant component.  Before the transition, finite $k$-clique components grow
and merge, so $\chi_{\rm finite}$ increases.  After the transition, much of
this mass has been transferred to the largest component and
$\chi_{\rm finite}$ decreases.  Its peak therefore marks the point at which
the finite sector is maximally loaded: large finite components are present,
but the largest component has not yet absorbed most of them.

The theoretical finite-sector transition coordinate is
\begin{equation}
q_\chi
=
\arg\max_q \mathbb E[\chi_{\rm finite}(q)] ,
\label{eq:q_chi_general}
\end{equation}
when the maximizer is unique; otherwise, $q_\chi$ denotes a chosen maximizer.
This is not a fitted crossing level; it is the maximum of the exact
finite-component susceptibility.

To express this object exactly, let $N_m(q)$ be the random number of
$k$-clique components with clique mass exactly $m$, and define
\begin{equation}
n_m(q)=\mathbb E[N_m(q)] .
\label{eq:n_m_definition}
\end{equation}
By linearity of expectation, the all-components second moment is
\(\mathbb E[\chi_{\rm all}(q)]=\sum_{m\ge1}m^2n_m(q)\).  The
largest-component subtraction requires the tail event in which at least one
component reaches the mass level \(s\).  With
\(Y_s(q)=\sum_{m\ge s}N_m(q)\), one has, for every finite graph,
\begin{equation}
\mathbb E[M_{\max}(q)^2]
=
\sum_{s\ge 1}(2s-1)
\bigl[1-\mathbb P(Y_s(q)=0)\bigr],
\label{eq:mmax_tail_void}
\end{equation}
and therefore
\begin{equation}
{
\mathbb E[\chi_{\rm finite}(q)]
=
\sum_{m\ge 1}m^2 n_m(q)
-
\sum_{s\ge 1}(2s-1)
\bigl[1-\mathbb P(Y_s(q)=0)\bigr].
}
\label{eq:exact_finite_sector_nm}
\end{equation}

All sums are finite in a finite graph because the number of possible
$k$-cliques is finite.  The exact ensemble object is therefore the pair
\(\bigl(n_m(q),\mathbb P(Y_s(q)=0)\bigr)\).  The first object gives the
second-moment mass of all components; the second subtracts the mass
already absorbed by the largest component. At its peak, finite-sector
production is equal to the absorption of the largest-component.  The corresponding
derivative condition is given in Appendix~\ref{app:component_count}. Thus, the finite-sector peak is the point where the production of finite-component second moment
is balanced by absorption into the largest component.

\subsection{Event-level coagulation interpretation}

The same balance can be read locally along any monotone graph process.
Suppose that a new \(k\)-clique is created and touches \(r\) distinct existing
\(k\)-clique components with clique masses \(m_1,\ldots,m_r\).  The full
second-moment increment contains the exact coalescence term
\(2\sum_{i<j}m_i m_j\): it appears only when the event connects at least two
preexisting components.  This is the point of contact with
Smoluchowski-type coagulation theory.  In classical coagulation models, a
coagulation kernel is prescribed; here the mass-product term follows directly
from the algebra of the second moment, while geometry and \(k\)-clique
adjacency decide which mergers are actually available.

If the new clique touches components of masses \(m_1,\ldots,m_r\), the merged
component has mass
\begin{equation}
m'=1+\sum_{i=1}^{r}m_i,
\label{eq:event_coag_mprime}
\end{equation}
and
\begin{equation}
\Delta\chi_{\rm all}
=
\left(1+\sum_{i=1}^{r}m_i\right)^2
-
\sum_{i=1}^{r}m_i^2
=
1+2\sum_{i=1}^{r}m_i
+
2\sum_{1\le i<j\le r}m_i m_j .
\label{eq:event_coag_delta_chi_all}
\end{equation}
The three terms correspond to birth of a new clique atom, one-component
growth, and true coalescence of preexisting components.  Since
\(\chi_{\rm finite}=\chi_{\rm all}-M_{\max}^2\), the corresponding
finite-sector increment is
\begin{equation}
\Delta\chi_{\rm finite}
=
\Delta\chi_{\rm all}
-
\left[(M_{\max}^{+})^2-(M_{\max}^{-})^2\right].
\label{eq:event_coag_delta_chi_finite}
\end{equation}
If an edge activation creates several \(k\)-cliques at the same value as
\(q\), one may list them in a fixed deterministic order and sum the elementary
increments; the total change in \(\chi_{\rm all}\) and
\(\chi_{\rm finite}\) is unchanged.

\subsection{Computation of the exact analytic component-count problem}

The preceding formulas reduce the macroscopic transition problem to two
objects: the component-count law $n_m(q)$ and the tail-void probabilities
$\mathbb P(Y_s(q)=0)$.  The first determines the full second moment; the
second determines how much of that moment is carried by the largest component.
Both objects have exact analytic representations by support integrals and
inclusion--exclusion over component-count factorial moments; the explicit
forms are given in the Appendix~\ref{app:component_count}.  This is the sense in
which the finite-sector construction is closed: no additional threshold rule is
missing.

The computational difficulty is now explicit.  For small $m$, one is
enumerating local motifs, and the support integrals involve small connected
patterns that can be evaluated or audited directly.  For large $m$, one is
enumerating possible finite clusters in geometry, together with their internal
clique connectivity, exposed boundary, overlapping exclusion regions, and
isolation from every outside vertex that could attach to them.  The support
must be connected in the $k$-clique space and have exactly the prescribed clique
mass. Thus, the bottleneck is the critical large-component tail problem
$m\gg1$, not the scalar branching criterion.

As a practical bridge between the exact formula and numerical evaluation, we
also use a one-large closure, described in Appendix~\ref{app:one_large}.  Its
role is only diagnostic: it asks how much of the largest-component subtraction
can be reconstructed from the mean component-count law alone.

\section{Hard-threshold random geometric graph}

We now specialize the general theory to the hard-threshold random geometric
graph.  The connection kernel is
\begin{equation}
W_R(r)=\mathbf 1_{\{r\le R\}},
\end{equation}
so, two vertices are connected exactly when their torus distance is not larger
than $R$.  This case is valuable because it is pure geometry: after the
positions are fixed, every edge, clique, completion lens, and exposed face is
determined by the radius $R$.

The natural control coordinate is the mean degree
\begin{equation}
q=(N-1)V_d^{\mathbb T}(R),
\end{equation}
where $V_d^{\mathbb T}(R)$ is the volume of a radius-$R$ ball on the unit
$d$-torus.  In the local Euclidean regime,
$V_d^{\mathbb T}(R)=v_dR^d$, and hence
\begin{equation}
R
\simeq
\left(\frac{q}{(N-1)v_d}\right)^{1/d}.
\end{equation}
Thus, fixed $q$ means that the radius of interaction shrinks as $N^{-1/d}$, while
the expected number of neighbors remains of order one.  This is the regime in
which the local geometric branching calculation has a nontrivial large-$N$
limit.

Let $Y=\{y_1,\ldots,y_{k-1}\}$ be a $(k-1)$-face reached during a local
$k$-clique exploration.  A new vertex completes this face if it lies in the
common intersection
\begin{equation}
L_Y(R)=\bigcap_{y\in Y}B_{\mathbb T}(y,R).
\end{equation}
In the local Euclidean regime, we write
$|L_Y(R)|=v_dR^d g_Y$, where $g_Y$ is the normalized lens volume.  The
conditional mean number of vertices that complete $Y$ is then
the finite-\(N\) binomial quantity
\begin{equation}
\lambda_Y^{(N)}
=
(N-k+1)|L_Y(R)|
=
q\,{N-k+1\over N-1}\,g_Y .
\end{equation}
After excess subtraction in
Eq.~\eqref{eq:Lambda_k}, the corresponding prefactor is
\((N-k)/(N-1)\). Thus, the large-\(N\), fixed-\(k\) form used below is
obtained by replacing this factor by one.  Substitution into the
size-biased continuation formula gives
\begin{equation}
\Lambda_k(q)
\simeq
q\,{N-k\over N-1}\,
\frac{\mathbb E[g_Y^2]}{\mathbb E[g_Y]}
=
q\,\frac{\mathbb E[g_Y^2]}{\mathbb E[g_Y]}
\left[1+O\!\left({k\over N}\right)\right].
\end{equation}
We denote the hard-kernel geometric factor by
\begin{equation}
H_{d,k}
=
\frac{\mathbb E[g_Y^2]}{\mathbb E[g_Y]}.
\end{equation}
Therefore,
\begin{equation}
B_k(q)\simeq (k-1)qH_{d,k},
\qquad
q_B\simeq \frac{1}{(k-1)H_{d,k}}.
\end{equation}

This formula is the hard-RGG analog of the classical local branching
criterion for Erd\H{o}s -- R\'enyi $k$-clique percolation.  The difference is
where the continuation probability comes from. In the Erd\H{o}s--R\'enyi graphs,
all candidate continuations are mixed through the whole set of vertex.  In the
hard RGG, the continuations pass through the geometric lens $L_Y(R)$, and the
exploration samples face a size bias proportional to their lens volume.
Thus, $B_k(q)=1$ is a genuine local onset: it detects when clique continuation
first becomes self-sustaining. However, it is not the full macroscopic
criterion for the formation of the percolating clique component.

The later component-level layer is described by the finite-component
susceptibility $\chi_{\rm finite}$ and its peak coordinate $q_\chi$, defined
in Eqs.~\eqref{eq:chi_finite_general} and~\eqref{eq:q_chi_general}.  In the
hard-threshold case these quantities keep the same meaning: they measure the
second-moment mass stored outside the largest $k$-clique component and locate
the point where this finite sector is maximally loaded.  This is a direct
analog of finite-cluster susceptibility in classical percolation theory.
The specialization is that the cluster size variable is clique mass rather
than vertex count.  A component may occupy a moderate number of vertices while
carrying many overlapping $k$-cliques.

For the hard kernel, the exact component-count problem also has a direct
geometric interpretation.  For each candidate finite support, one can, in
principle, determine the internal clique-adjacency structure, the clique mass
$m$, the attachable $(k-1)$-faces, the external attachment region from which an
outside vertex would connect to the component, and the corresponding
no-attachment probability. Thus, the finite-sector object is not an unknown
phenomenological rule: it is an explicit geometric enumeration of isolated
finite $k$-clique components.  The formal support integral is given in
Appendix~\ref{app:component_count}.

The important point is the remaining problem of finite-component enumeration.
For small components, this enumeration is local. However, for a large component clique
mass $m$, it must sum over many possible spatial shapes,
clique-adjacency structures, exposed boundaries, and overlapping exclusion
regions. Thus, the large-$m$ computation is not a simple extension of the
scalar criterion $B_k=1$; it is a constrained finite-component enumeration
problem.

\subsection{Separation of local and finite-sector criticalities in hard-threshold RGG}

\begin{figure}[H]
\centering
\includegraphics[width=\textwidth]{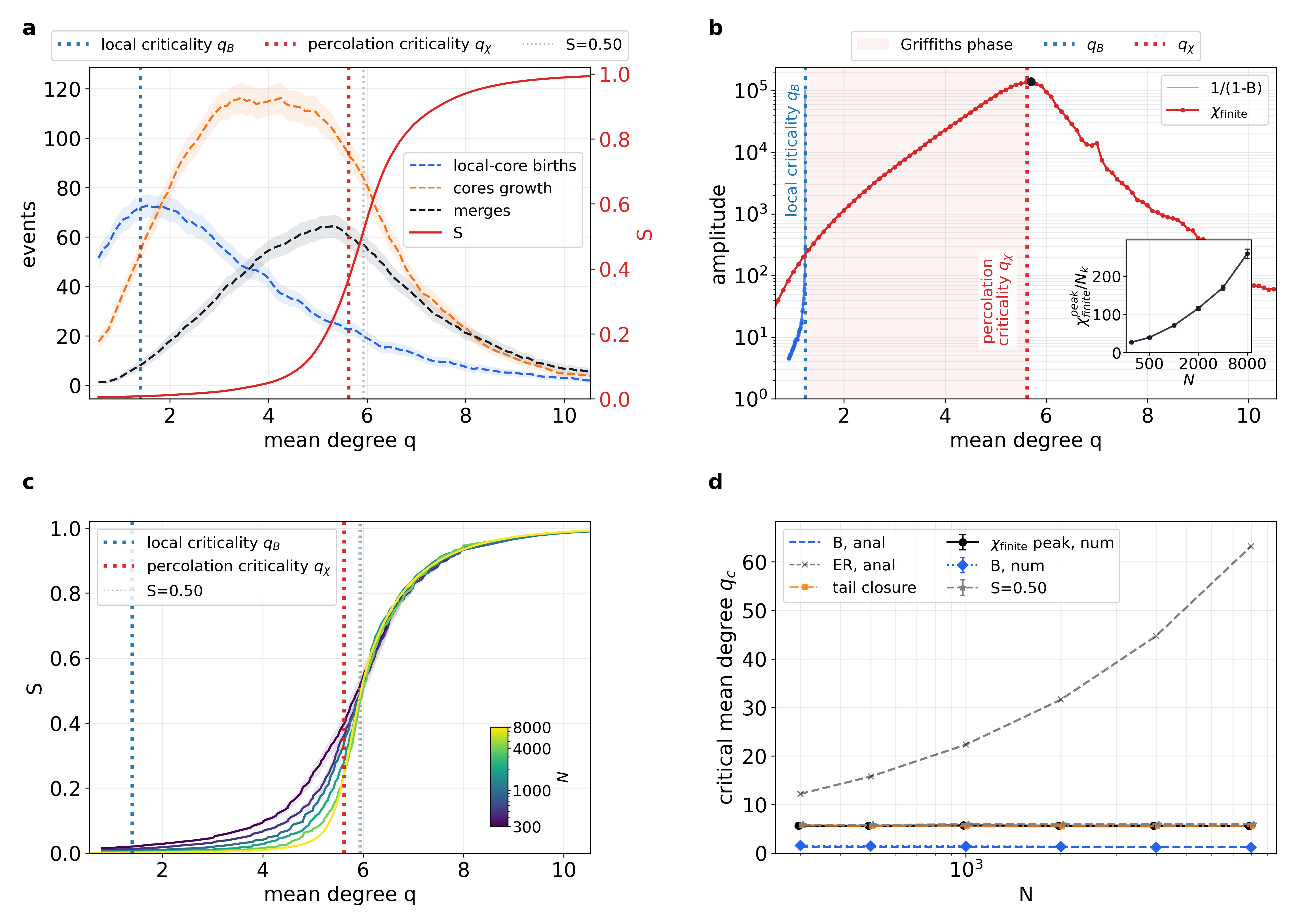}
\caption{Percolation of \(k\)-cliques in a hard-threshold random geometric
graph. The elementary
event rates separate local-core births, growth of
existing triangle components, and mergers of distinct triangle components;
the red curve on the right axis is the largest-component fraction $S$
(\(N=1000,\ d=4,\ k=3,\ n_{\rm stat}=100\)).  Panel b
places the two critical response objects on the same $q$ axis: the fine-grid
local response $(1-B(q))^{-1}$ diverges at the local branching onset,
whereas the finite-sector susceptibility $\chi_{\rm finite}$ peaks later when
finite-component production is balanced by absorption into the largest
component (\(N=1000,\ d=4,\ k=3\); \(n_{\rm stat}=50\) for the local
response, \(n_{\rm stat}=200\) for the finite-sector response, and
\(n_{\rm stat}=100\) for its tail extension).  Panels c and d test the same coordinates under system-size
scaling: panel c shows $S(q)$, while panel d compares the extracted critical
mean degrees $q_c$ for $B$ analytic, $B$ numerical, tail closure,
$\chi_{\rm finite}$ peak, $S=0.50$, and the Erd\H{o}s--R\'enyi analytic
reference (\(d=4,\ k=3,\ n_{\rm stat}=100\), with
\(n_{\rm stat}=200\) for the high-\(N\) extension).}
\label{fig:fin_hard_1}
\end{figure}

Figure~\ref{fig:fin_hard_1} shows the emergence of the percolating
$k$-clique component as a structural phase transition in the hard-threshold
RGG.

Panel a separates three elementary event types.  The isolated local-core
birth curve counts events that create new finite triangle components.  The
one-component growth curve counts events in which a new triangle attaches to
one existing triangle component and increases its mass.  The component-merger
curve counts events in which a new triangle connects two or more existing
components.  The red curve is the largest vertex fraction $S$, which records
the macroscopic percolation transition.

The separation between the vertical local-onset marker and the increase in $S$ is
the first key observation.  The system does not jump directly from the local
self-sustaining clique continuation to a macroscopic $k$-clique component.
Instead, local continuation first creates spatially localized finite
components.  These components grow inside geometric neighborhoods, then begin
to merge through exposed faces, and only later does the largest component
absorb a macroscopic fraction of the graph.  The fixed level $S=0.50$ is shown
as a visual midpoint of the largest-component transition, not as the
definition of the transition.

Panel b shows the corresponding divergent response variables.  The quantity
$B(q)$ is the mean number of new clique continuations generated by one clique already
reached through its remaining forward faces.  When $B<1$, repeated
local continuation has a finite expected response,
\begin{equation}
1+B+B^2+\cdots=\frac{1}{1-B}.
\end{equation}
Thus, $(1-B)^{-1}$ is not merely a convenient plotting transform; it is the
local branching susceptibility to clique continuation, and it diverges at the
local onset $B=1$.  The finite-sector response is instead the component-level
susceptibility $\chi_{\rm finite}$, whose maximum occurs later.  The inset
shows the same effect after normalization by the total number of cliques
$N_k$: the unnormalized peak grows extensively and even
$\chi_{\rm finite}/N_k$ continues to increase in the displayed system sizes.
The second criticality is therefore an explosion of the finite-sector second
moment, not only an extensive-counting effect.  In physical terms, the
characteristic finite-component mass stored outside the largest component
becomes system-size dependent before that mass is absorbed by the waterfall.

Panels c and d test this interpretation under system-size scaling.  Panel c
shows the sharpening of the largest-component waterfall with $N$; panel d then
extracts the critical coordinates.  The numerical $B=1$ points follow the
analytic small-ball prediction, so $B$ does more than mark an early crossing:
it carries the fundamental local scaling law of clique continuation.  Once
this local critical law is separated from the later finite-sector peak
$ q_\chi$, the system can occupy a whole interval in which local
continuation is already critical but global clique percolation has not yet
occurred.  This is the structural opening for the Griffiths phase discussed
in the following.  The finite-sector peak, the conditional one-large reconstruction, and
the fixed $S=0.50$ reference lie later because they involve the growth and
absorption of finite components.  The Erd\H{o}s--R\'enyi reference emphasizes
the contrast: geometric packing and spatial correlations allow the
percolating clique component to form at much lower mean degree, while also
separating local continuation from the later component-level transition.

The process-level picture, therefore, separates two singular responses.
The first is the local branching susceptibility at $q_B$, where
self-sustaining clique continuation becomes possible and the response
$(1-B)^{-1}$ diverges.  The second is the finite-sector susceptibility at
$q_\chi$, where the population of finite $k$-clique components is finally
absorbed into the percolating component.  In the hard-threshold RGG these two
points enclose an extended intermediate regime:
\begin{equation}
q_B<q<q_\chi.
\end{equation}
We identify this intermediate regime as a Griffiths phase: local
supercritical pockets exist, but global clique percolation is delayed by
geometric localization and by the slow coalescence of finite components.

\subsection[Genealogy of the Griffiths phase]{Structure of the Griffiths phase}

\begin{figure}[H]
\centering
\includegraphics[width=\textwidth]{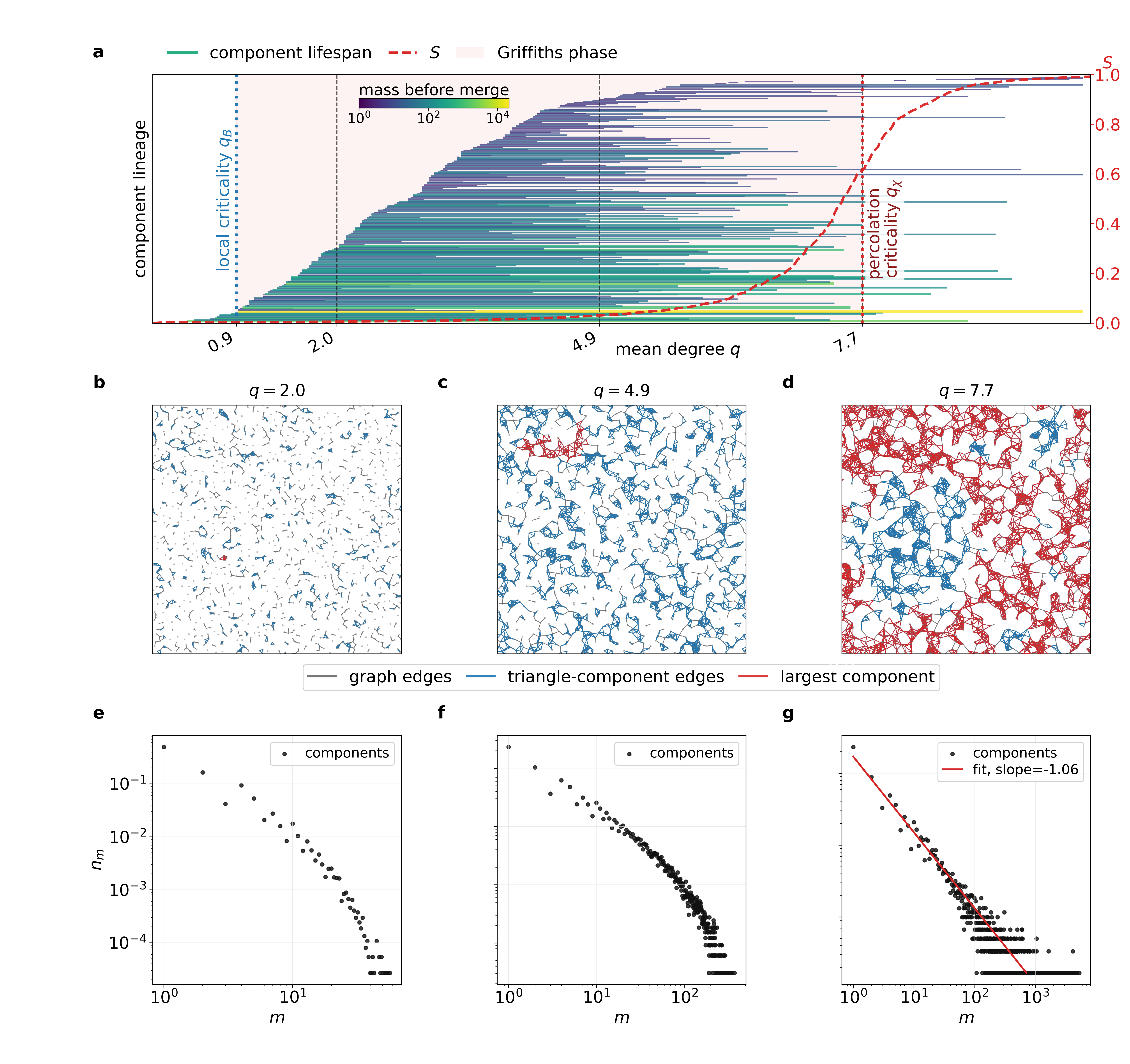}
\caption{Genealogy of the Griffiths phase in a two-dimensional
hard-threshold RGG.  Panel a shows lifetimes
of finite triangle-component lineages whose local component-level branching
has become supercritical; the shaded interval $q_B<q<q_\chi$ marks the
Griffiths phase, and the dashed red curve is the ensemble largest-component
waterfall $S(q)$.  Panels b--d show graph snapshots at $q=2.0$, at the
midpoint between this value and $q_\chi$, and at $q_\chi$; background edges
show the geometric graph, blue edges mark triangle-component edges, and red
edges mark the largest triangle component.  Panels e--g show the
ensemble-mean finite-component mass distributions at the same
three $q$ values.  The distributions broaden across the Griffiths phase and
approach their cleanest scale-free form near $q_\chi$
(\(N=2000,\ d=2,\ k=3,\ n_{\rm stat}=150\)).}
\label{fig:hard_step_2d_genealogy}
\end{figure}

Figure~\ref{fig:hard_step_2d_genealogy} visualizes the same separation in
a single two-dimensional hard-threshold realization, while keeping the
 waterfall and component-mass distributions of the ensemble as statistical
references.  Its purpose is to make the Griffiths phase visible as a regime of
long-lived locally active clique components, not only as a gap between two
threshold markers.

After the first criticality, local supercritical components begin to
nucleate.  They do not immediately become the giant clique component. Instead
, they survive over a broad interval in $q$, grow by local clique completion,
and slowly coalesce through attachable $(k-1)$-faces.  The waterfall occurs only
when this finite-sector population has accumulated enough mass and enough
contacts that component mergers and absorption into the largest component
dominate.

The genealogy plot is therefore the real-space counterpart of the
susceptibility separation in Fig.~\ref{fig:fin_hard_1}.  It shows why the
intermediate regime should be described as a phase rather than as a narrow
finite-size rounding: locally active finite components persist, interact, and
develop a broad mass distribution throughout a long interval before the final
percolation event.  Across this interval, the finite-sector ensemble already
contains a persistent power-law tail and a broad realization-to-realization
spread of component masses.  What changes near $q_\chi$ is not the sudden
birth of the tail, but its maturation into the cleanest scale-free regime: the cutoff moves
outward, mesoscopic components become common, and the finite-sector
susceptibility becomes dominated by the same broad tail that is about to be
absorbed into the largest component.

\subsection{Scale-free finite-component statistics at the transition}

\begin{figure}[H]
\centering
\includegraphics[width=\textwidth]{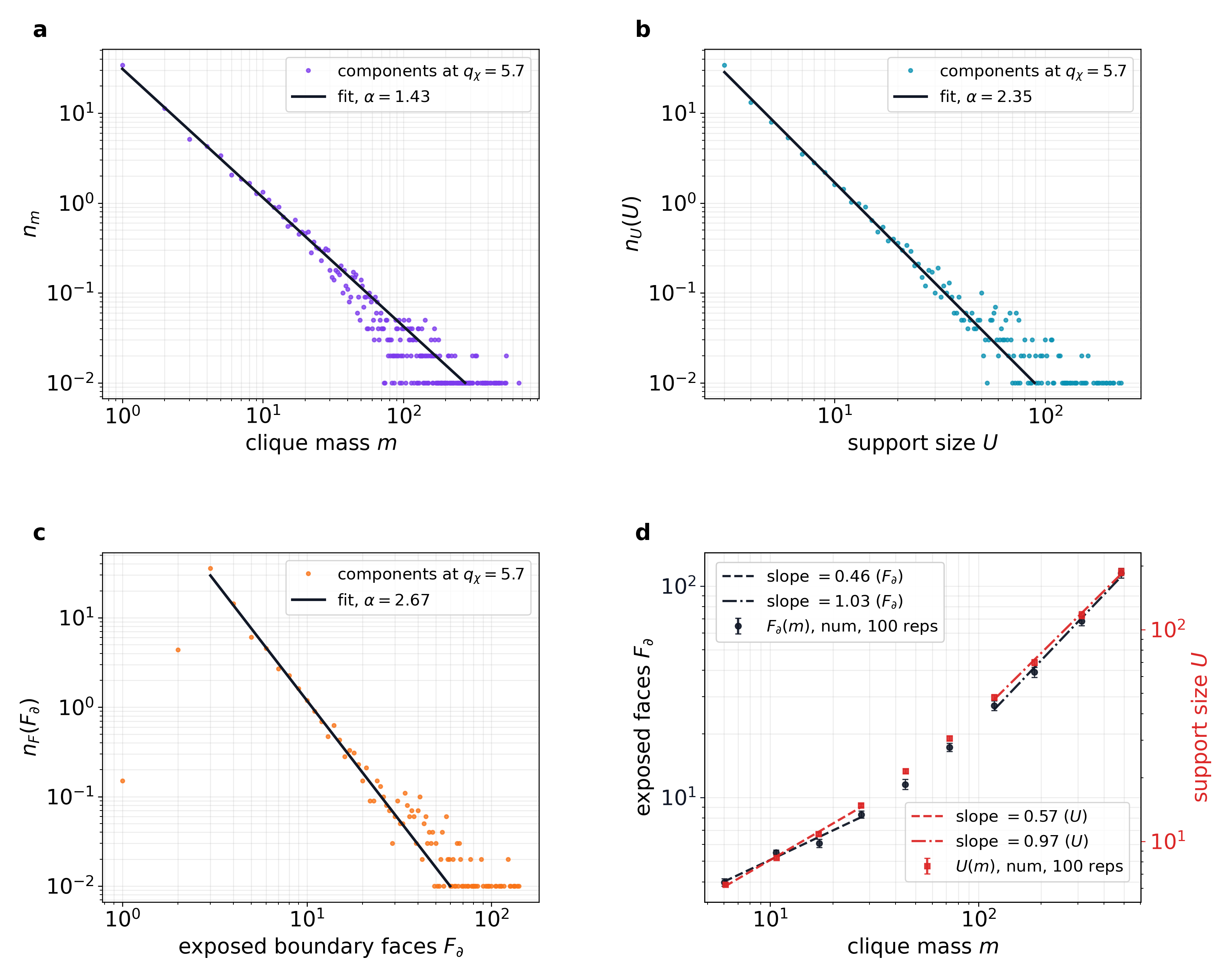}
\caption{Scale-free finite-sector statistics at the hard-threshold transition.
The finite-sector peak is at $ q_\chi=5.7$.  Panels a--c are log-log component-count laws at
$q_\chi$.  Panel a shows the clique-mass distribution $n_m(m)$,
where $m$ is the number of $k$-cliques in a finite component; the displayed
body fit gives the mass exponent $\alpha\simeq1.43$.  Panel b shows the
support-size distribution $n_U(U)$, where $U$ is the number of vertices in the
component vertex support; the body fit gives $\alpha\simeq2.35$.  Panel c
shows the distribution $n_F(F_\partial)$ of exposed boundary faces, fitted on the
positive-boundary body; components with $F_\partial=0$ are omitted from the
log-log plot.  Panel d relates the same objects morphologically: exposed
boundary $F_\partial(m)$ is plotted on the left axis and vertex-support size
$U(m)$ on the right axis as functions of clique mass $m$, with separate
early-body and late-body log-log slopes.  These slopes are diagnostics of compact local
filling versus extended rough-frontier growth, not additional threshold
definitions.  The broad finite-component tail begins already inside the Griffiths phase and reaches
its cleanest scale-free form near $q_\chi$
(\(N=1000,\ d=4,\ k=3,\ n_{\rm stat}=100\)).}
\label{fig:fin_hard_2}
\end{figure}

Figure~\ref{fig:fin_hard_2} focuses on the finite-component distribution
itself, which becomes more scale-free near the critical growth stage
of the percolating component.  This is the part of the theory that remains
computationally hard in a fully analytic evaluation.  The exact formal object
is known: $n_m(q;N,d,k)$ and the corresponding tail void probabilities determine
$\mathbb E[\chi_{\rm finite}(q)]$.  What is difficult is the large-$m$
evaluation of this component-count law without first sampling graphs.

Panel a shows the mean component-count law $n_m(m)$, defined by
Eq.~\eqref{eq:n_m_definition}, at the finite-sector peak.
The broad log-log body is well described by a power-law form over the
resolved finite-size window.  This is the finite-component analog of the
Fisher cluster-size picture in ordinary percolation: near criticality, the
finite-cluster distribution loses a single characteristic size and develops a
scale-free regime, while finite size and off-criticality still provide
cutoffs.  The exponent should not be interpreted as an Erd\H{o}s--R\'enyi
universal number.  It is a geometric clique-component exponent for the
particular hard-RGG ensemble and the mass variable.

Together with Fig.~\ref{fig:hard_step_2d_genealogy}, this panel
shows that the finite-component tail is not created only at the instant of the
waterfall.  The Griffiths phase already contains long-lived finite components
with a widening mass distribution; $q_\chi$ is the point where this broad
finite-sector population is maximally susceptible to absorption into the
largest component.

Panels b and c show that the same loss of a characteristic scale is visible
in two geometric projections of the finite components. The size of the support $U$
counts the number of vertices in the support of the vertex of the component, while the exposed
boundary $F_\partial$ counts the open $(k-1)$-faces through which the
component can still grow or merge.  The broad laws in $U$ and $F_\partial$
therefore show that the scale-free behavior in $n_m$ is not just an
artifact of counting overlapping triangles. Numerically, these objects are
consistent with each other: the mass, support, and boundary are different
projections of the same critical finite components, not three independent
mechanisms.

Panel d makes the relation between mass, support, and boundary explicit.  The
exposed boundary $F_\partial(m)$ and the vertex-support size $U(m)$ are plotted
against the mass of the component clique and fitted separately in two local log-log
windows.  The early-body fits describe small and moderately sized components,
where growth is dominated by compact local filling.  The late-body fits
describe larger finite components, where the objects become rougher and the
exposed frontier grows almost proportionally to the vertex support.  These
fits are not competing threshold definitions; they are a morphology
diagnostic.  They show how the volume-like clique mass is coupled to a
surface-like set of open growth channels through which finite components
merge before being absorbed into the largest component.

Together, Fig.~\ref{fig:fin_hard_2} explains the computational bottleneck of
the exact hard-RGG theory.  A fully analytic prediction of $q_\chi$ requires
the large-$m$ component-count law $n_m(q)$ and the corresponding largest-tail
void probabilities.  Near the transition this law is broad and nearly
scale-free, so the problem is not reducible to a few small component
integrals.  This scale-free window should be read as positive evidence of the
theory rather than as a failure of closure.  The formal closure points
precisely to $n_m(q)$ and the tail-void probabilities; the data show that,
near the percolation transition, these objects enter a critical finite-size
regime with no single characteristic component mass.  That is exactly the
regime where one should expect a separate effective theory of critical
clique clusters, rather than a simple continuation of the local calculation.

\subsection{Dimension and clique-size scaling}

\begin{figure}[H]
\centering
\includegraphics[width=0.95\textwidth]{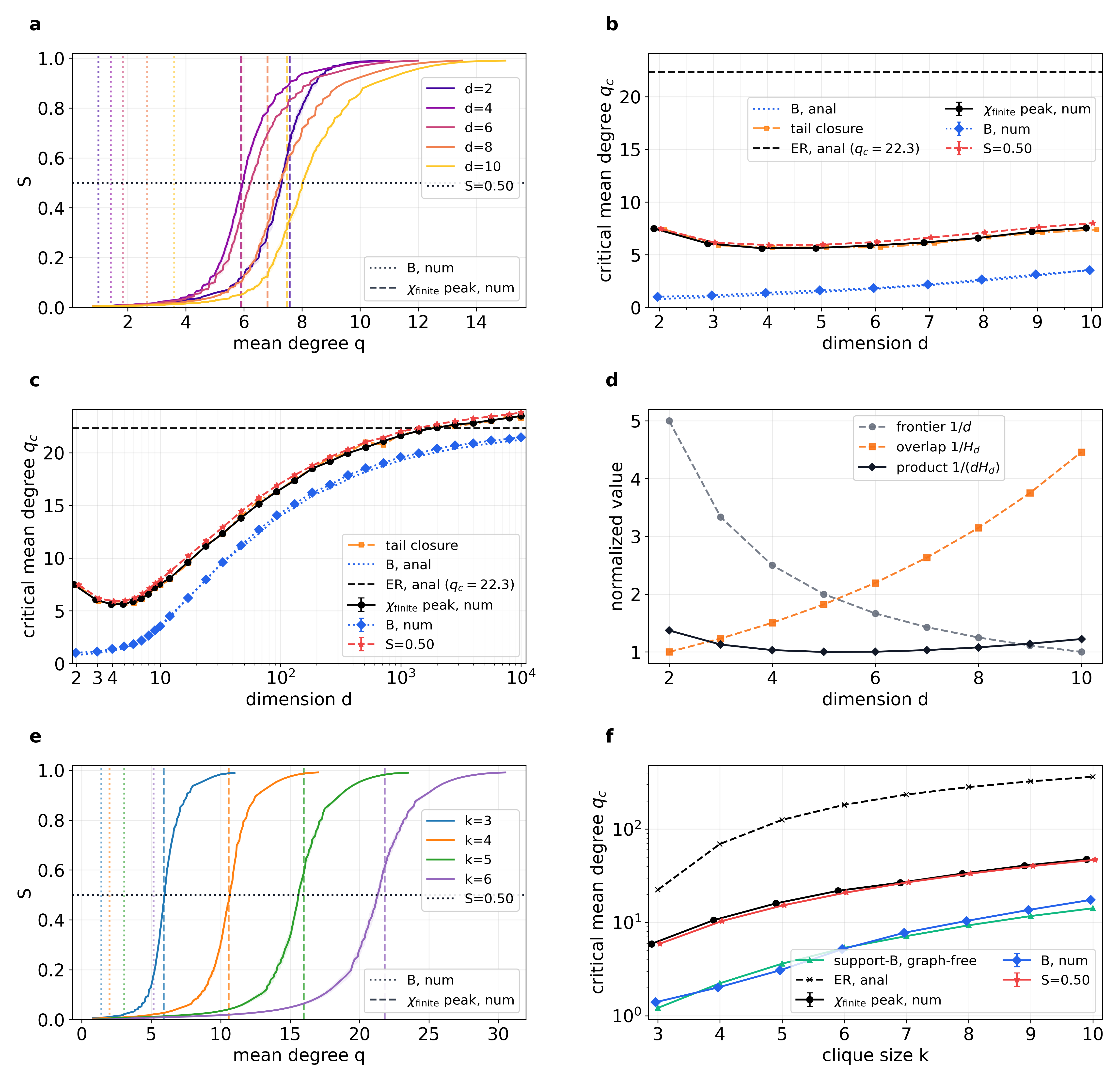}
\caption{Dimension and clique-size scaling for the hard-threshold RGG.  Panel a shows
largest-component curves $S(q)$ for $d=2,4,6,8,10$; vertical markers indicate
the analytic local onset $B$ anal, the numerical local onset $B$ num, and the
numerical $\chi_{\rm finite}$ peak, while the horizontal dotted line marks
$S=0.50$ (\(N=1000,\ k=3,\ n_{\rm stat}=100\)). 
%The panel shows that the gap between the local onset and the
%formation of the percolating clique component changes with dimension.
Panel b
summarizes the low-dimensional critical mean degrees $q_c$ for the same
threshold coordinates: $B$ anal, $B$ num, tail closure, numerical
$\chi_{\rm finite}$ peak, fixed $S=0.50$, and the Erd\H{o}s--R\'enyi analytic
reference (\(N=1000,\ k=3,\ n_{\rm stat}=100\)).  Panel c extends this dimension scan to the displayed high-$d$
range; the hard-RGG finite-sector coordinates approach the fully mixed
reference only gradually, while the local $B$ scale remains a separate
geometric onset (\(N=1000,\ k=3,\ n_{\rm stat}=100\)).  Panel d gives an analytic mechanism-level decomposition of the
low-dimensional U-shape using different normalized geometric factors.  Panel e shows $S(q)$ for
$k=3,4,5,6$ with vertical markers for $B$ anal, $B$ num, and
$\chi_{\rm finite}$ peak; larger $k$ shifts the percolating component to
higher mean degree because each continuation requires a higher-order
simultaneous completion (\(N=1000,\ d=4,\ n_{\rm stat}=100\)).  Panel f compares the corresponding critical mean
degrees as functions of $k$: the black curve is the numerical
$\chi_{\rm finite}$ peak, the blue curve is $B$ num, the red curve is
$S=0.50$, the black dashed curve is the Erd\H{o}s--R\'enyi analytic reference,
and the green analytic support-$B$ curve is a support-stability diagnostic
for reinforced $(k-1)$-faces rather than a new universal threshold
(\(N=1000,\ d=4\); \(n_{\rm stat}=100\) for \(k=3,\ldots,6\) and
\(n_{\rm stat}=60\) for \(k=7,\ldots,10\)).}
\label{fig:fin_hard_3}
\end{figure}

Figure~\ref{fig:fin_hard_3} examines how the finite-sector transition changes
with increasing dimension and clique size.  The central point is not only the
existence of a robust extended phase between the local onset and the formation
of the percolating clique component.  Importantly, the gap between these
scales and the finite-sector threshold itself varies non-monotonically with
dimension.  The low-dimensional curve has a U-shaped profile, suggesting a
competition between two geometric resources: a surface-like frontier through
which finite clique components find merger channels and a dimension-dependent completion penalty
coming from the shrinking overlap of high-dimensional lenses at fixed mean
degree.

Panel a shows that changing \(d\) does not shift the largest-component
transition and the local \(B\) onset in parallel.  The central observation in
panels b and c is the low-dimensional minimum: the finite-sector peak, the
one-large reconstruction, and the fixed-\(S\) reference form a U-shaped curve,
whereas the local \(B\) onset follows a different geometric scale. The
Erd\H{o}s--R\'enyi curve is only the fully mixed comparison.  The hard-RGG
threshold is lower because geometric packing helps, but the location of the
finite-sector transition is controlled by spatial merger geometry.

Panel d gives the geometric interpretation of the U-shape: the frontier
factor $1/d$ represents the surface-like merger channel, the overlap factor
$1/H_d$ represents the lens-completion penalty, and their product
$1/(dH_d)$ captures the competition between having more frontier directions
and making simultaneous face completion harder. Increasing
dimension gives a finite component more independent frontier directions through
which it can meet other components: finite components have more exposed
\((k-1)\)-faces through which they can find neighboring components and merge.
This frontier is the clique-percolation analog of a surface.  At the same
time, simultaneous completion of a \((k-1)\)-face becomes more selective,
because the relevant intersection lens becomes harder to occupy.  A vertex
completing a face reached must lie in a high-dimensional overlap region, while
the interaction radius at fixed mean degree follows the dimension-dependent
scaling $R\simeq(q/((N-1)v_d))^{1/d}$.  The observed threshold minimum comes
from the competition between these two effects: surface-like merger
opportunities initially help the finite sector connect, whereas
high-dimensional lens overlap eventually makes clique continuation harder.  The
observed U-shape is the finite-sector signature of this competition between a
surface-controlled law of component merger and a volume-like law of internal
clique accumulation and simultaneous face completion.

This also explains why the boundary variable $F_\partial$ in
Fig.~\ref{fig:fin_hard_2} is physically central.  It is not merely another
projection of the size of the components.  It counts the exposed growth and merger
channels through which Griffiths-phase components find each other before the
finite-sector susceptibility peak.

The panels e and f vary the size of the clique at fixed $d=4$.  As $k$ increases, the
largest-component transition shifts to larger $q$ because a continuation must
satisfy a higher-order simultaneous clique-completion constraint.  Panel f
shows the same separation of layers: the analytic $B$ curve is a
local-geometry baseline, the numerical $B$ curve measures the local onset,
and the $\chi_{\rm finite}$ peak and fixed-$S$ reference describe the delayed
finite-sector transition.  For $k>3$, the curve $B_{\rm loc}=1$ should be read
as a conditional local onset: a stable finite-graph crossing also requires
enough reinforced $(k-1)$-faces, measured by
$\mathcal R_2=\mathbb E\sum_Y C_Y(C_Y-1)$.  For larger cliques, the apparent local
boundary can therefore also be controlled by the point at which the graph has
enough link density to create a stable population of reinforced faces at all.
The measured $\mathcal R_2$ curve uses graph runs, whereas the analytic
$\mathcal R_2$ curve uses local geometry; this layer separates the formal local
branching condition from the additional support-density requirement needed for
high-order clique continuation.  The Erd\H{o}s--R\'enyi curve is included as
a fully mixed reference and grows differently with $k$.

The four hard-threshold figures therefore support a single mechanism.  The
local onset $B_k(q_B)=1$ is fundamental and captures the correct geometric
scaling of local clique continuation, while the observable clique-percolation
transition is controlled by the finite-component susceptibility peak $q_\chi$
defined in Eq.~\eqref{eq:q_chi_general}.  Between these two scales, finite
$k$-clique components grow, expose frontiers, merge through surface channels,
become approximately scale-free in the critical window, and are finally
absorbed by the largest component.  This is the hard-kernel
Griffiths-like interval. Its non-monotone U-shaped dependence on dimension
shows that the finite-sector transition is controlled by a competition between
surface-like merger channels and the dimension-dependent cost of simultaneous
face completion.

The work therefore gives a complete analytic decomposition of the transition
problem: $B_k(q_B)=1$ gives the local continuation onset, while $q_\chi$ is
determined by the finite-component susceptibility.  The remaining hard object
is not an unknown threshold rule, nor a matter of insufficient sampling
statistics.  It is the critical large-$m$ law of finite clique clusters,
$n_m(q)$ for $m\gg 1$, and the associated tail-void probabilities.  The
numerical evidence shows that this finite-sector law carries a critical
finite-size structure: an extended power-law tail, a cutoff front, and
mesoscopic components that are ultimately absorbed by the largest cluster.  A
full analytic description would require the distribution of large finite
components together with the cutoff  of the tail, their shape, exposed boundary,
and isolation probabilities.
This is a separate effective theory of critical $k$-clique clusters in
geometric space.  The present theory localizes
this open problem in a mathematically specific and physically interpretable
way.

\section{Power-law kernel}
\label{sec:powerlaw_radial_scale_hierarchy}

\subsection{Radial scale hierarchy}

We now turn from the hard-threshold kernel to the power-law kernel
\begin{equation}
W_\alpha(r)=\min\{1,(r_0/r)^\alpha\}.
\label{eq:powerlaw_kernel}
\end{equation}
The local and finite-sector percolation objects have already been defined.
What changes here is the radial scale structure of the kernel itself.  A
power-law tail does not reduce to one geometric scale: the probability of a
long connection decays, but the number of possible geometric placements grows
with radius.  The resulting balance creates several scale regimes.  Moreover,
the relevant balance is not the same for edges and for larger cliques,
because a closure event contains several kernel factors.

We first describe the radial supply of possible events before imposing the
percolation dynamics.  For a \(k\)-clique continuation, the closure event
contains \(k-1\) new kernel factors, and the corresponding radial shell measure
is
\begin{equation}
\mathrm d\mu_k(r)\propto r^{d-1}W_\alpha(r)^{k-1}\,\mathrm dr .
\label{eq:powerlaw_shell_measure_r}
\end{equation}
Here \(r^{d-1}\mathrm dr\) is the geometric shell volume, while
\(W_\alpha(r)^{k-1}\) is the power-law suppression from the \(k-1\)
simultaneous connections needed to continue a reached \((k-1)\)-face.  Edges
are the one-factor case, while a 3-clique continuation requires two new kernel
factors to close a triangle.  This count is not the static
number \(\binom{k}{2}\) of edges in a clique; it is the dynamical number of
new kernel factors required by a continuation step.  The hierarchy to
observe is therefore the hierarchy in clique continuation size.

From the same radial measure we use the log-shell density
\begin{equation}
\psi_k(r)=\frac{\mathrm d\mu_k}{\mathrm d\log r}
\label{eq:powerlaw_psi_definition}
\end{equation}
which counts event weight per equal multiplicative radial shell, for example
, from \(r\) to \(br\).  This is the natural scale density for a power-law
kernel, because multiplying \(r\) by a constant changes the kernel by a fixed
factor.  In the tail \(r>r_0\),
\begin{equation}
\psi_k(r)\propto r^{\zeta_k},
\qquad
\zeta_k=d-(k-1)\alpha .
\label{eq:powerlaw_zeta_definition}
\end{equation}
Thus, \(\zeta_k=0\) is a plateau per logarithmic radial shell.  A density per
equal additive radial increment would define a different shifted marginality 
\(d-1-(k-1)\alpha=0\); below the scale analysis is formulated only in
logarithmic shells.

\begin{figure}[H]
\centering
\includegraphics[width=\textwidth]{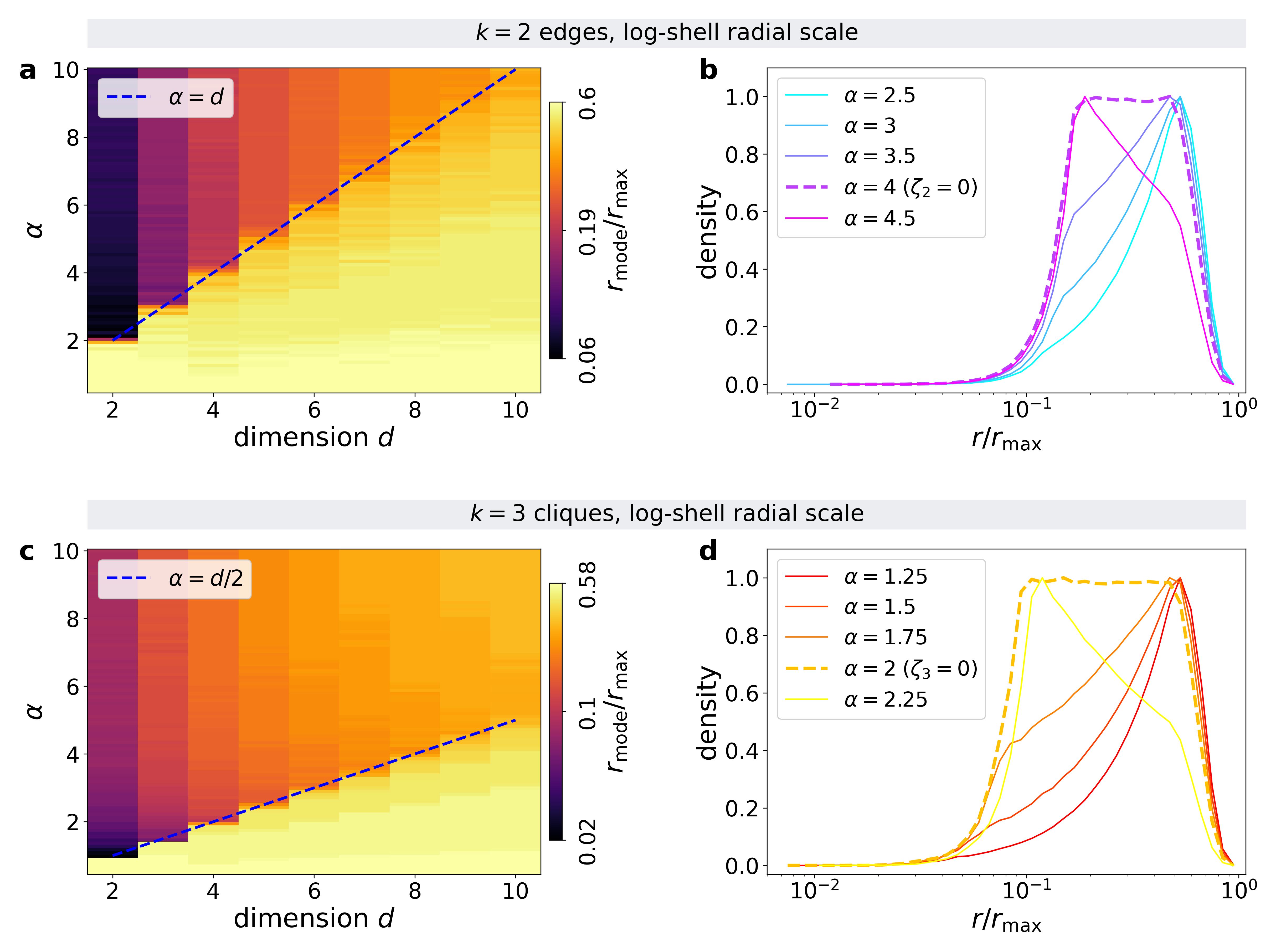}
\caption{Radial modes and log-shell profile densities for the power-law
kernel.  Panels a and c show the mode location
\(r_{\rm mode}/r_{\max}\) in the \((d,\alpha)\) plane for edges and
3-cliques (\(N=500,\ n_{\rm stat}=5\)).  Panels b and d show the
corresponding representative profile cuts for the log-shell density
\(\psi_k=\mathrm d\mu_k/\mathrm d\log r\)
(\(N=500,\ d=4,\ n_{\rm stat}=100\)).  Its plateau occurs at
\(\zeta_k=d-(k-1)\alpha=0\), where equal logarithmic shells carry equal bare
event weight.  The mode maps show that this balance occurs at different
values of \(\eta=\alpha/d\) for different clique sizes.}
\label{fig:powlaw-modes-profiles}
\end{figure}

Figure~\ref{fig:powlaw-modes-profiles} shows why the power-law case cannot be
summarized by a single locality boundary.  In panels a and c the color gives
the radial position of the dominant mode, and one sees several geometric
subregimes separated by the predicted log-shell marginal lines.  The relevant
profile question is which multiplicative shell carries the event weight.
Panels b and d show the \(d=4\) log-shell profile cuts and the plateau at
\(\zeta_k=0\):
near this condition no logarithmic shell is preferred, while away from it the
weight is pulled toward either the core or the outer scale.  With
\(x=\log(r/r_0)\), the shell measure therefore takes the log-scale form
\begin{equation}
\mathrm d\mu_k(x)\propto e^{\zeta_k x}\,\mathrm dx .
\label{eq:powerlaw_zeta_measure_x}
\end{equation}
At this point, the logarithmic scale structure is visible directly in the
measure.  The coordinate \(x=\log(r/r_0)\) transforms the multiplicative changes in
the radius into additive scale steps, so the logarithmic radial shells become the
natural shells of a multiplicative tail.  The sign of \(\zeta_k\)
classifies the scale regime:
\begin{equation}
\zeta_k>0:\ {\rm outer\ scale\ weighted},\qquad
\zeta_k=0:\ {\rm log\ radial\ marginality},\qquad
\zeta_k<0:\ {\rm core\ weighted}.
\label{eq:powerlaw_zeta_signs}
\end{equation}
Thus, \(\zeta_k=0\) is a scale-marginality condition, not a percolation
threshold.  It says that equal logarithmic shells carry equal bare event
weight.

\subsection{Clique-size dependent hierarchy of boundaries between the effective ER and geometric regime}
\label{subsec:powerlaw_clique_size_scale_hierarchy}

The log-radial shell measure itself explains why clique continuations of different
sizes form a radial scale hierarchy.  A step in \(x=\log(r/r_0)\) corresponds to a
multiplicative change in physical radius, and the coefficient \(\zeta_k\) is
the signed growth rate of the \(k\)-clique continuation weight per logarithmic
scale step.
Therefore, it is useful to introduce the normalized exponent
\(\eta=\alpha/d\).  This is the natural coordinate because the competition is
between \((k-1)\alpha\), the kernel suppression accumulated by
the \(k-1\) new connections, and
\(d\), the growth rate of the logarithmic shell volume.  In the \((d,\alpha)\)
mode maps of Fig.~\ref{fig:powlaw-modes-profiles}a,c, the marginal lines are
straight rays; in \(\eta=\alpha/d\) they become fixed
clique-size boundaries.
The  \(d=4\) cuts  in Fig.~\ref{fig:powlaw-modes-profiles}b,d make the same fact
visible as plateau formation in the log-shell density.

The normalized exponent rewrites the log-shell slope as
\(\zeta_k=d[1-(k-1)\eta]\).  The scale boundary for a
\(k\)-clique continuation is
therefore
\begin{equation}
\eta_k^\ast=\frac{1}{k-1},\qquad k\ge2.
\label{eq:powerlaw_eta_boundary}
\end{equation}
Different clique-continuation sizes cross their boundaries at different
values of \(\eta\).  Larger cliques become local first; edges can
remain long-range over the same parameter range.  For the 3-clique
problem, edges are the one-factor case with boundary \(\eta=1\),
while 3-clique closure requires two new connections and has boundary
\(\eta=1/2\).  The three basic sectors for 3-cliques are
\begin{equation}
\begin{array}{lll}
\eta<1/2
&:& \text{edges and 3-cliques are long-range and ER-like},\\[1mm]
1/2<\eta<1
&:& \text{mixed regime: edges remain long-range, closures localize},\\[1mm]
\eta>1
&:& \text{local geometric regime already at the edge level}.
\end{array}
\label{eq:powerlaw_k3_regime_example}
\end{equation}
This ordered set of clique-size boundaries replaces a single
long-range/local crossover.

\begin{figure}[H]
\centering
\includegraphics[width=\textwidth]{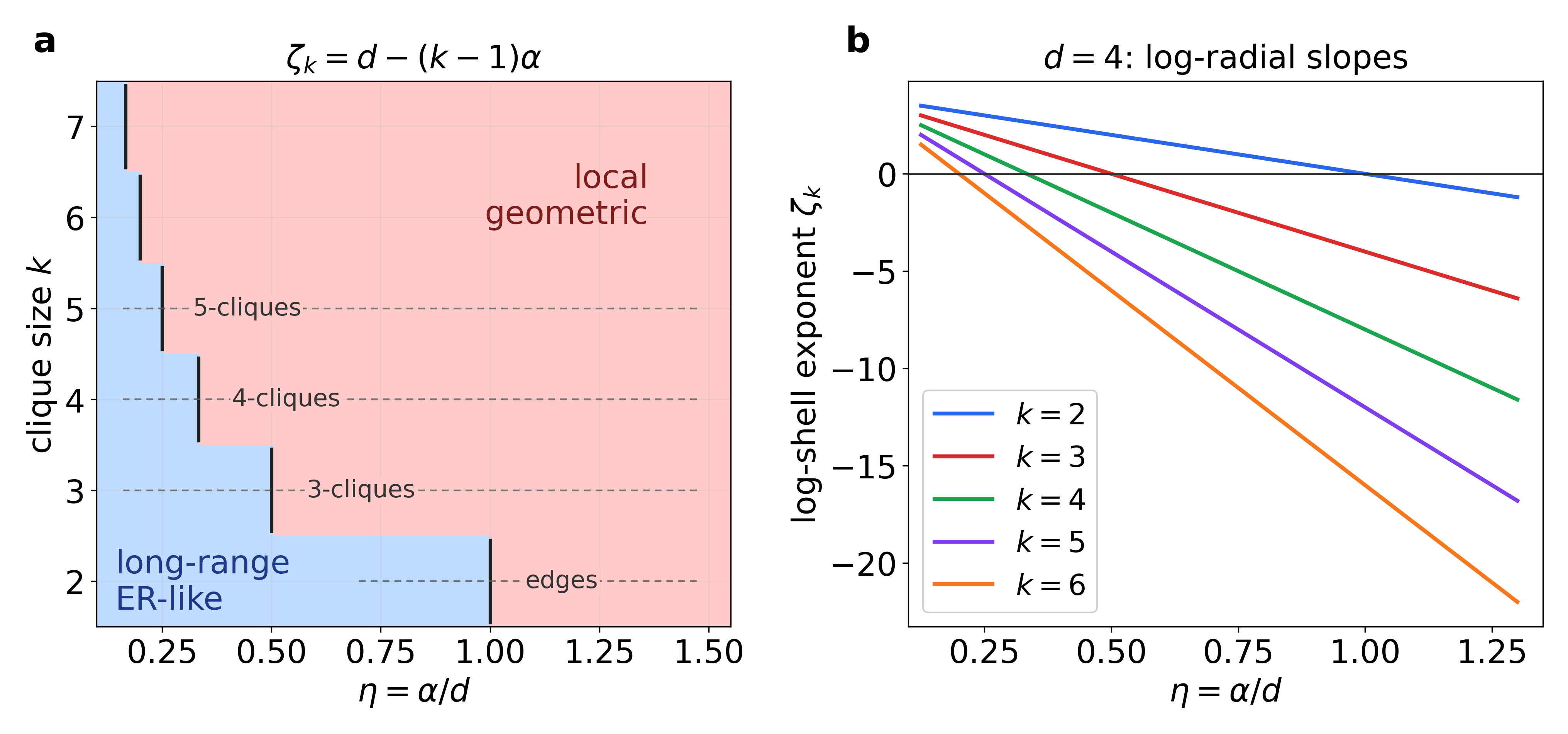}
\caption{Log-shell clique-size slope hierarchy.  Panel a is the analytic
regime map obtained directly from the marginal condition
\(\zeta_k=d-(k-1)\alpha=0\), equivalently
\(\eta=1/(k-1)\), for \(k=2,\ldots,7\), i.e. one through six
new kernel factors.  Blue regions are long-range or ER-like at the
level of bare scale supply, while red regions are local/geometric.  Panel b
shows measured log-shell slopes: for each
clique size, the
solid curve crosses zero at the corresponding \(\zeta_k=0\) boundary.
Positive slopes mean that outer logarithmic shells carry more weight; negative
slopes mean that the weight is pulled toward the core
(\(N=2000,\ d=4,\ n_{\rm stat}=50\)).}
\label{fig:powlaw-slope-hierarchy}
\end{figure}

Figure~\ref{fig:powlaw-slope-hierarchy} displays the
clique-size log-shell law
directly.  Panel a is the scale hierarchy used below: a \(k\)-clique
continuation loses long-range relevance at \(\eta=1/(k-1)\).
Panel b is a
direct slope measurement at fixed \(d=4\).  Its vertical axis is the fitted
power of the measured log-shell density
\(\psi_k(r)\sim r^{\zeta_k}\).
Zero on this axis therefore means a plateau per logarithmic shell; positive
values mean that larger scales carry more weight, and negative values mean
that the weight is pulled toward the core.  This is a useful slope
for a power-law kernel because it compares equal multiplicative scale shells.

\begin{figure}[H]
\centering
\includegraphics[width=\textwidth]{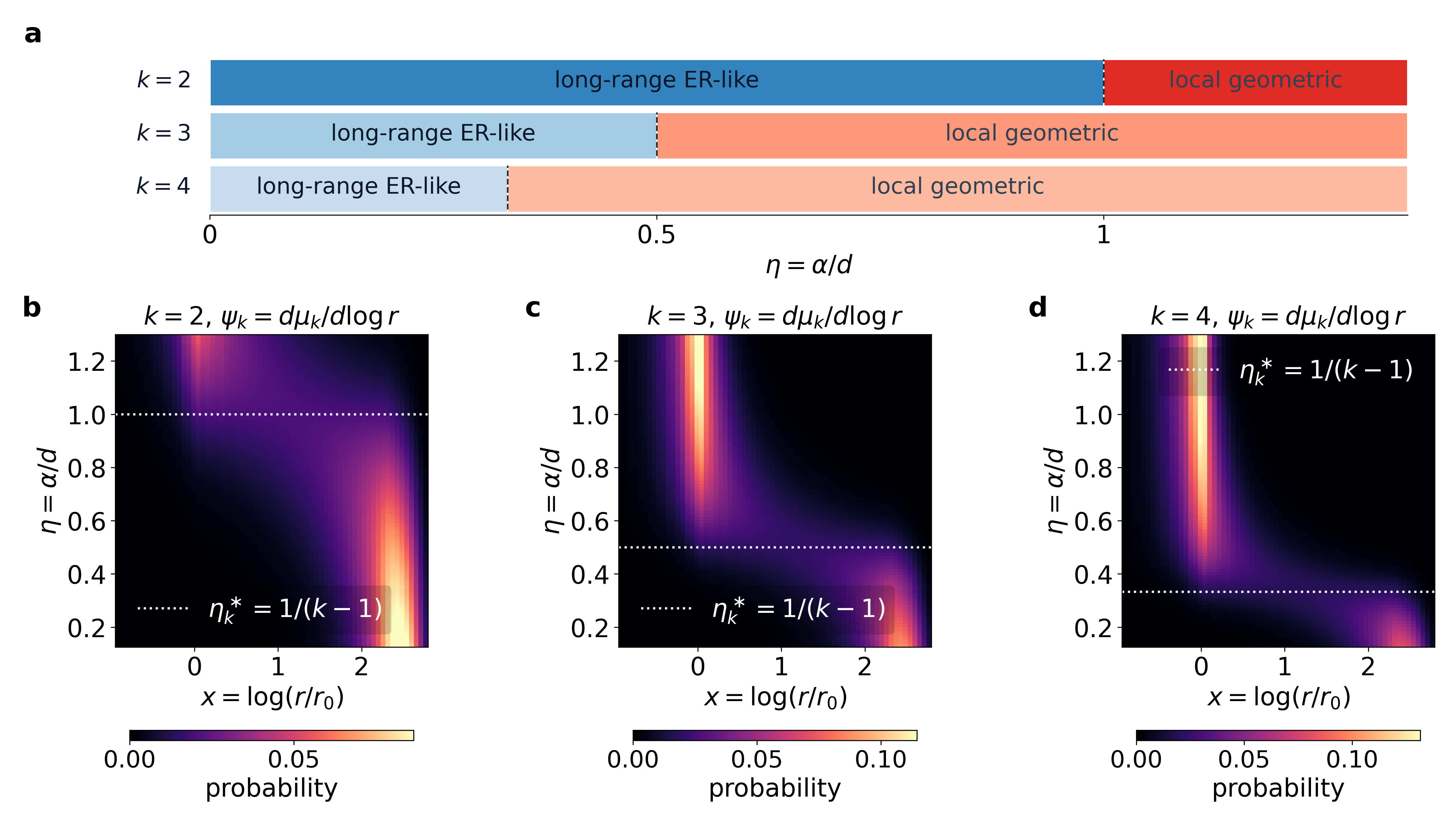}
\caption{Bare log-radial scale distributions for the edge,
3-clique, and 4-clique continuations.
Panel a is an analytic regime table in \(\eta=\alpha/d\): for each clique size it marks
the long-range side and the local side of the log-shell boundary
\(\eta=1/(k-1)\).  Panels b--d show the row-normalized log-shell density
\(\psi_k=\mathrm d\mu_k/\mathrm d\log r\) as a function of scale
\(x=\log(r/r_0)\) and \(\eta\) for one-, two-, and three-factor
continuation events (\(N=2000,\ d=4,\ n_{\rm stat}=50\)).  Near log-shell marginality,
the heatmaps become approximately uniform across \(x\), meaning that no
single logarithmic shell dominates the bare continuation scale supply.  The
color bar reports the normalized probability within each \(\eta\)-row.}
\label{fig:powlaw-regime-ladder}
\end{figure}

Figure~\ref{fig:powlaw-regime-ladder} displays the clique-size hierarchy
as actual scale distributions.  Panel a shows the ordered locality cascade:
4-cliques cross their log-shell boundary before 3-cliques, and edges
cross last. Panels b--d show the corresponding
row-normalized scale densities for one-, two-, and three-factor continuation events.
For a fixed \(\eta\), the color
indicates where the available clique events live inside the radial
scale window.  Near \(\eta=1/(k-1)\), the density spreads across almost the entire
\(x\)-axis, so no single logarithmic shell dominates the bare event supply.
Away from these marginal lines, the same density concentrates either near the
core or near the outer scale.  This visible widening of the scale distribution
is the geometric source of the mixed power-law regimes.

The sign of \(\zeta_k\) gives the asymptotic regime, but a finite graph sees
only a finite logarithmic radial interval.  Let this interval be
\begin{equation}
0\le x\le \ell,
\qquad
\ell=\log(r_{\max}/r_0),
\label{eq:powerlaw_finite_window}
\end{equation}
where \(r_{\max}\) is the largest radial scale accessible to the observable in
the torus.  The important finite-window scale coordinate is
\begin{equation}
X_k=\zeta_k\ell .
\label{eq:powerlaw_Xk}
\end{equation}
This coordinate is important because it combines two pieces that cannot be
separated in a finite system: \(\zeta_k\), the signed growth or decay rate of
the \(k\)-clique continuation shell weight per unit logarithmic scale and \(\ell\), the
number of logarithmic scale units available in the finite window.  Their
product is the total logarithmic bias accumulated between the inner and outer
ends of the window.  A weak nonzero slope can therefore become visible across a
long window, while a stronger slope can have little effect if the available
window is short. Thus, \(X_k\) is the finite-window coordinate for the
clique-size scale hierarchy.

The sign of \(X_k\) has the following direct meaning:
\begin{equation}
\begin{array}{lll}
X_k>0 & \Longrightarrow &
\text{the finite window is weighted toward large-radius shells},\\[2mm]
X_k=0 & \Longrightarrow &
\text{logarithmic shells are balanced and the scale law is marginal},\\[2mm]
X_k<0 & \Longrightarrow &
\text{the finite window is weighted toward the core}.
\end{array}
\label{eq:powerlaw_X_interpretation}
\end{equation}
Here \(x\) is only the internal logarithmic radius used to define the window;
the regime variable is \(X_k\).  In particular,
\begin{equation}
X_2=d(1-\eta)\ell,\qquad
X_3=d(1-2\eta)\ell,\qquad
X_4=d(1-3\eta)\ell
\label{eq:powerlaw_X234}
\end{equation}
organize edge supply, 3-clique closure supply, and 4-clique continuation
supply.  For \(k=3\), the
transition to \(k\)-clique percolation in the described geometry is controlled
primarily by \(X_2\) and \(X_3\).  If these coordinates have different signs or
very different magnitudes, edges and closures draw their available events from
different radial sectors.  This is the finite-window form of the mixed regime.

Thus, the power-law kernel produces a clique-continuation hierarchy of
locality.  The same value of \(\eta\) can place edges, 3-clique closures, and
higher-order continuations in different radial regimes, and the natural
finite-system coordinates for this hierarchy are precisely the corresponding
\(X_k\) coordinates.

\subsection{\(k\)-clique percolation and the Griffiths phase in the power-law scale hierarchy}
\label{sec:powerlaw_percolation_scale_hierarchy}

The local onset \(q_B\), the finite-sector peak \(q_\chi\), the
largest-component curve \(S(q)\), and the finite-sector susceptibility
\(\chi_{\rm finite}(q)\) are now placed in the scale hierarchy derived above.
The local continuation layer depends on the radial supply of closure events,
while the finite-sector layer is governed by how the resulting components
accumulate, persist, and merge across the available scale sectors.

\begin{figure}[H]
\centering
\includegraphics[width=0.85\textwidth]{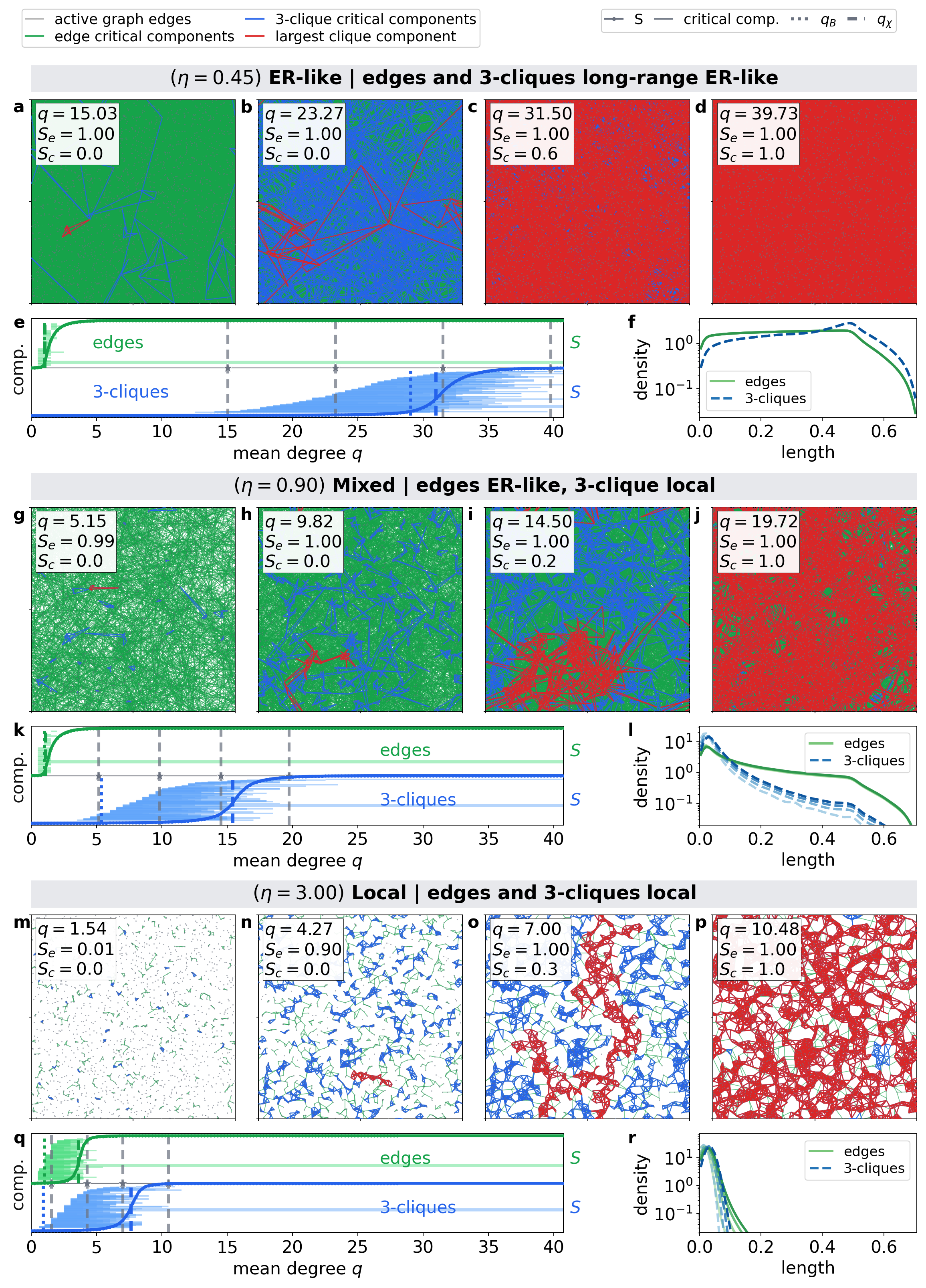}
\caption{$\eta$-scale storyboard shows the same \(q\)-evolution through three different radial-scale
regimes.  
%  From top to bottom the rows move from the ER-like nonlocal regime
%\(\eta=0.45\), where both edges and 3-clique closures are long-range, to the
%mixed regime \(\eta=0.90\), where edges remain broad but 3-clique closures are
%already localizing, and finally to the local geometric regime \(\eta=3.00\),
%where both edges and 3-cliques are local. 
Within each row the horizontal direction
is the evolution in mean degree \(q\): the four graph snapshots
\((a\)--\(d\), \(g\)--\(j\), \(m\)--\(p)\) show representative states through
the same response window, while the wide waterfall--lifeline panel
\((e,k,q)\) places those states on the curves \(S_{\rm edge}(q)\) and \(S(q)\)
together with the local onset \(q_B\) and the finite-response guide
\(q_\chi\).
The scale-distribution panels \((f,l,r)\) give the radial mechanism behind the
snapshots: they compare the active edge-length distribution, associated
with \(k=2\), with the 3-clique closure scale,
associated with \(k=3\) and measured by
the triangle diameter
\(r_{\max}(t)=\max(r_{ij},r_{il},r_{jl})\) 
%Broad distributions indicate
%ER-like nonlocal placement; concentration at short distances indicates
%geometric localization.  Green denotes edge-critical connectivity, while blue
%and red denote 3-clique components, with red marking the largest 3-clique component
(\(N=2000,\ d=2,\ k=3,\ n_{\rm stat}=100\)).}
\label{fig:powlaw-eta-channel-storyboard}
\end{figure}

Figure~\ref{fig:powlaw-eta-channel-storyboard} places percolation on top of the
scale hierarchy by treating edges and 3-clique closures as different
clique sizes.  The rows, from top to bottom, show the ER-like nonlocal regime, the
mixed regime, and the fully local geometric regime.  The row titles already
give the finite-window
coordinates: at \(\eta=0.45\), \(X_2>0\) and \(X_3>0\), so both edges and
3-clique closures are long-range and ER-like; at \(\eta=0.90\), \(X_2>0\)
but \(X_3<0\), so edges remain ER-like while 3-clique closures are already
local; at \(\eta=3.00\), both coordinates are negative, and both edges and
3-cliques are local.
Each row then follows the evolution in \(q\), and the graph snapshots show how
the same scale regimes are dynamically expressed as \(q\) increases near
the Griffiths-like response window.

In the ER-like row, panels a--d, both edges and 3-cliques percolate
through nonlocal long-range connections, and the percolation law of both is
close to the ER-like regime.  The Griffiths-like
interval is not absent in this row; it is strongly compressed in \(q\), so the
local onset and finite-sector response are nearly simultaneous, as in
Erd\H{o}s--R\'enyi graphs.  Panel e
nevertheless shows the characteristic ER-like early activity: many small
3-clique components can appear and survive over a visible \(q\)-interval, but
they are mostly isolated triangle events with little continuation structure.
They do not collectively reproduce the next layer of 3-clique connectivity, so
the network-averaged branching criterion reaches \(B=1\) only later.  Panel f
confirms the same interpretation in scale space: active edge lengths and
3-clique diameters are broad and sample-dependent rather than organized into a
coherent localized critical cluster family.

In the mixed row, panels g--j, the contrast is sharper and is the clearest
visual example of the clique-size scale hierarchy.  Panel k shows edge
connectivity already present on the ER-like background while 3-cliques
develop localized critical pockets.  Panel l shows why: edge lengths
still occupy broad nonlocal scales, but the 3-clique closure scale has already
moved toward short geometric distances.  This is the literal situation \(X_2>0\),
\(X_3<0\): the locality appears for 3-cliques before it appears for
the edges.  In the local row, panels m--p, edges, and 3-cliques grow through
geometric neighborhoods.  Panel q then shows the corresponding
lifeline behavior, while panel r shows both scale distributions concentrated
at short distances. The largest red 3-clique component expands by spatial
coalescence rather than by nearly mean-field placement of long-range
connections.

\begin{figure}[H]
\centering
\includegraphics[width=\textwidth]{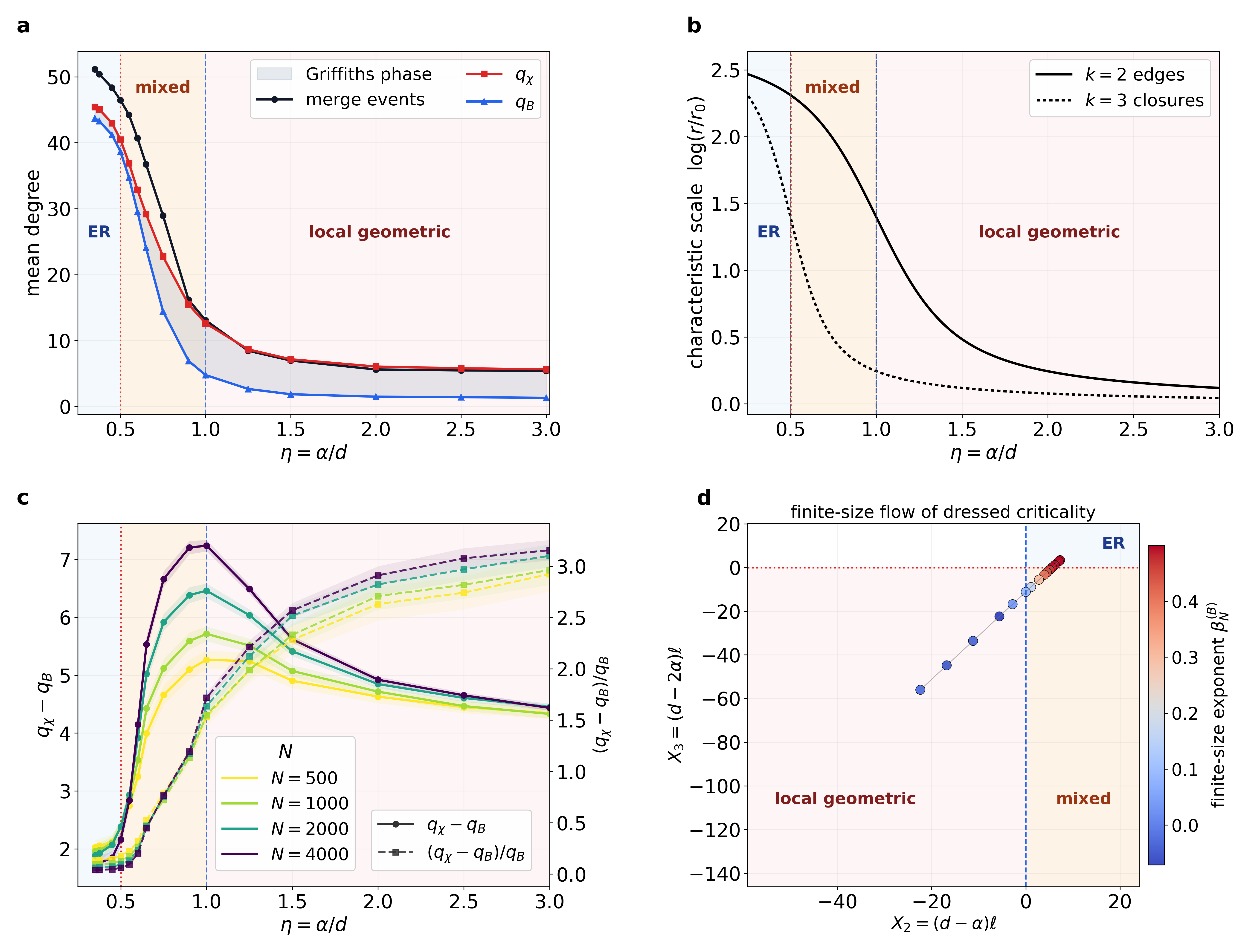}
\caption{Power-law thresholds, characteristic scales, Griffiths-like width,
and finite-size drift for \(k=3\).  Panel a plots threshold coordinates versus
\(\eta=\alpha/d\), including the merge-event peak \(q_{\rm merge}\), the
finite-sector peak \(q_\chi\), and the local onset \(q_B\)
(\(N=4000,\ d=4,\ k=3\); \(n_{\rm stat}=50\), with
\(n_{\rm stat}=75\) or \(100\) at the top-up points).  Panel b analytically plots the
characteristic log-shell coordinate \(\langle\log(r/r_0)\rangle\) for edges
\(k=2\) and 3-cliques \(k=3\); the vertical
guides \(\eta=1/2\) and \(\eta=1\) are the closure and edge log-shell
marginalities.  Panel c shows the absolute width \(q_\chi-q_B\) and the
relative delay \((q_\chi-q_B)/q_B\), equivalently
\(q_\chi/q_B=1+(q_\chi-q_B)/q_B\), across
\(\eta\) for several system sizes \(N\)
(\(d=4,\ k=3,\ n_{\rm stat}=50\)).  Panel d replots the finite-size drift
\(\beta_N^{(B)}\) of the local onset in the scale coordinates \((X_2,X_3)\),
where \(X_2\) is the edge scale coordinate and \(X_3\) is the 3-clique closure
scale coordinate (\(d=4,\ k=3,\ n_{\rm stat}=49\)--\(200\)).}
\label{fig:powlaw-width-waterfalls}
\end{figure}

Having seen the clique-size mechanism, Fig.~\ref{fig:powlaw-width-waterfalls}
summarizes the same effect on \(\eta\), \(N\), and the finite-window scale
coordinates.  Panel a shows the threshold curves: it separates local onset
\(q_B\), finite-sector response \(q_\chi\), and merge activity as functions of
the power-law exponent.  The merge curve adds an important diagnostic.  In the
ER-like regime, the merge-event peak occurs substantially to the right of both
\(q_B\) and \(q_\chi\): local and finite-sector responses have already
appeared before the main merger activity, as expected when long nonlocal
connections assemble components in a nearly mean-field manner.  In the local
geometric regime, by contrast, the merge peak moves onto the finite-sector
response scale. There, the second critical landmark is controlled by spatial
coalescence of already local 3-clique clusters.

Panel b gives the scale-level reason for the mixed window: the edge scale
remains extended after the 3-clique closure scale has already moved toward the
local side.  Panel c measures the resulting Griffiths-like delay directly.  In
the ER-like nonlocal regime the delay is not zero in principle, but it is
minimal: the two percolation landmarks are very close because both
clique sizes are
effectively broad and mean-field-like.  In this sense, the ER-like
Griffiths-like interval is almost instantaneous.  In the fully local regime, the
delay is clear but comparatively stable, because both are governed by
local geometry.  The largest and most size-sensitive width appears in the
mixed regime, where edge supply and 3-clique closure are controlled by
different scale sectors.

Panel d is the compact scale-coordinate reading of this behavior.  The color
\(\beta_N^{(B)}\) measures how the local onset \(q_B\) drifts with the size of the system.
Positive \(\beta_N^{(B)}\) means that the local-onset scale grows with system
size, while values near zero indicate a size-stable local threshold.
Plotted in \((X_2,X_3)\), the drift changes character near the marginal
axes:
\(X_3=0\) is the 3-clique closure boundary and \(X_2=0\) is the edge boundary.
This is why the peak of the delay is most naturally read in the \(X\)-plane.
The separation \(q_\chi-q_B\) is largest where the edges and 3-cliques are
not described by the same radial scale ensemble.

\begin{figure}[H]
\centering
\includegraphics[width=\textwidth]{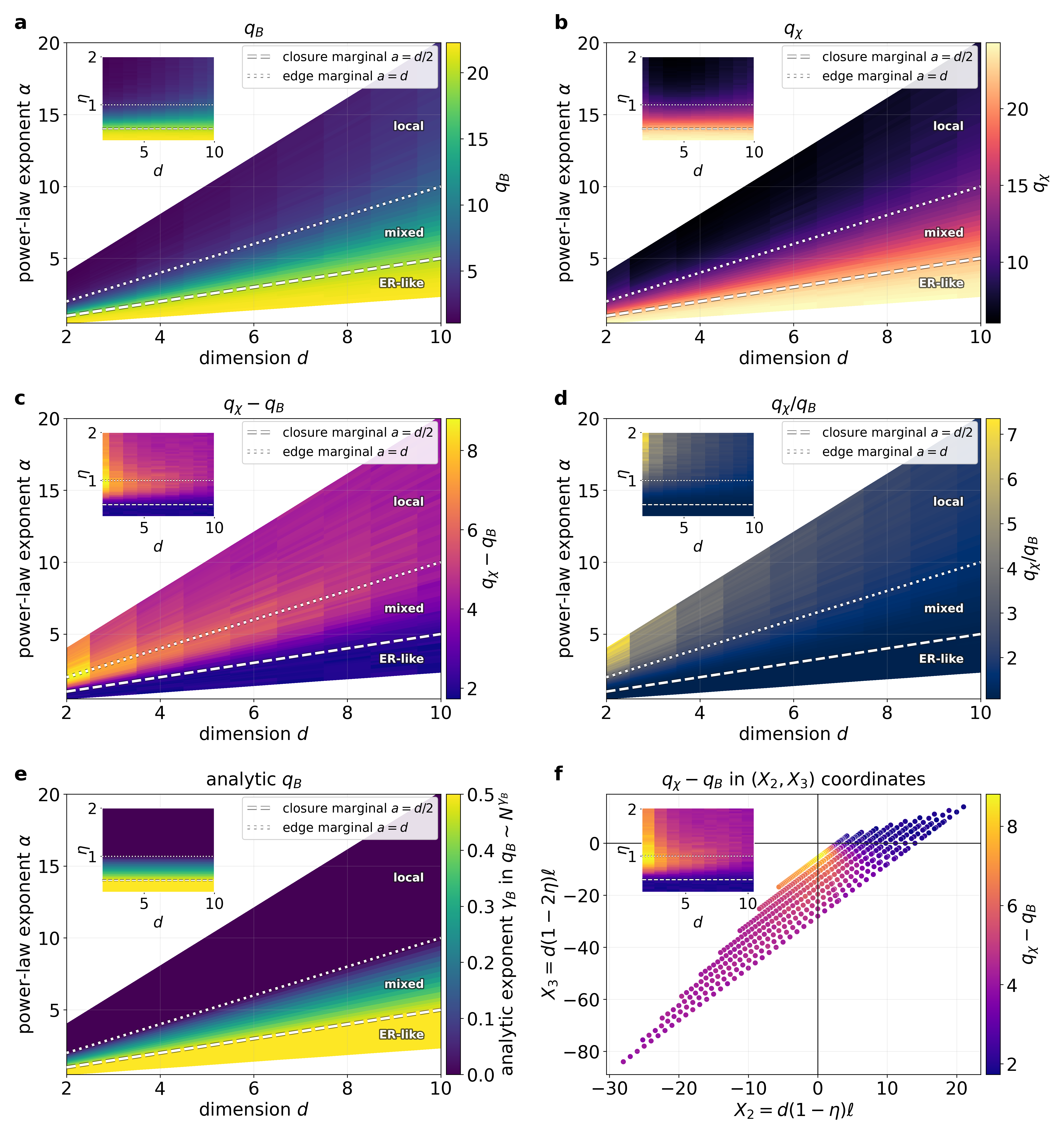}
\caption{Power-law percolation map and scale-coordinate bridge for
\(k=3\).  Panels a--d show numerical
threshold and delay surfaces in the physical \((d,\alpha)\) plane:
\(q_B\), \(q_\chi\), \(q_\chi-q_B\), and
\(R_G=q_\chi/q_B\).  The insets replot the same quantities in
\((d,\eta)\), and the white guides mark the closure and edge log-shell
marginalities \(\alpha=d/2\) and \(\alpha=d\)
(\(N=1000,\ 2\le d\le10,\ n_{\rm stat}=50\)).  Panel e shows the analytic
local-onset exponent \(\gamma_B\) in \(q_B\sim N^{\gamma_B}\).  Panel f is the
scale-coordinate map: each sampled point is plotted in
\((X_2,X_3)\), where \(X_2=d(1-\eta)\ell\) is the edge coordinate and
\(X_3=d(1-2\eta)\ell\) is the triangle-closure coordinate, and the color is
the Griffiths-like width \(q_\chi-q_B\).}
\label{fig:powlaw-alpha-d-scale-map}
\end{figure}

Figure~\ref{fig:powlaw-alpha-d-scale-map} places the numerical
percolation observables first in the physical kernel plane \((d,\alpha)\).  Panels
a and b separate local onset from finite-sector response; panels c and d show
the same separation as absolute and relative delay.  The insets are important:
they show that the apparent \((d,\alpha)\) structure is not arbitrary but is
organized by the dimensionless coordinate \(\eta=\alpha/d\).

The analytic scaling of the local onset follows from the shell integral 
\(J_k(r_0)=\int r^{d-1}W_\alpha(r)^{k-1}\mathrm dr\) for a \(k\)-clique
continuation.  The derivation is given in
Appendix~\ref{app:powerlaw_shell_integrals}.  For \(k=3\), where closure
requires two new kernel factors, the resulting local-onset scale is
\begin{equation}
q_B=
\begin{cases}
N^{1/2}, & \eta<1/2,\\
(N/\log N)^{1/2}, & \eta=1/2,\\
N^{1-\eta}, & 1/2<\eta<1,\\
\log N, & \eta=1,\\
O(1), & \eta>1 .
\end{cases}
\label{eq:powerlaw_qB_scaling_k3}
\end{equation}
Thus, the same clique-size boundaries that organize the bare radial measures
also organize the analytic local-onset map.  Panel e shows this statement in
heatmap form: the exponent changes at the closure boundary, the edge
boundary, and the logarithmic marginal lines.

Panel f is the scale-coordinate map \((X_2,X_3)\).  In raw
\((d,\alpha)\) coordinates, the maximum of \(q_\chi-q_B\) is visible, but its meaning
is partly obscured by the simultaneous variation of dimension, kernel exponent,
and finite scale window.  In the \(X\)-coordinates, the maximum is located near
the crossing sector of the marginal axes \(X_2=0\) and \(X_3=0\), where
edges and 3-cliques change scale character inside the same finite-window
coordinate system.  This is the main link between the bare scale hierarchy and
the dressed Griffiths-like response.  The peak of \(q_\chi-q_B\) is not
only a feature in the \(q\)-space; it appears where the edge coordinate and the
3-clique closure coordinate approach their marginal sectors.  This is the
natural place for a large delay: edge-scale opportunities can still reach
outward, whereas closure events are close to marginality or are becoming
core-weighted.  The power-law Griffiths-like width is therefore the dressed
percolation response to a mixed clique-size scale regime.

The same \(X_k\) coordinate can now be written as a normalized finite-window
scale ensemble.  With \(x=\log(r/r_0)\), on \(0\le x\le\ell\), the normalized
density of the bare \(k\)-clique continuation scale is
\begin{equation}
p_k(x)=\frac{e^{\zeta_k x}}{Z_k(\ell)},
\qquad
Z_k(\ell)=\int_0^\ell e^{\zeta_k x}\,\mathrm dx .
\label{eq:powerlaw_finite_window_density}
\end{equation}
Thus, \(Z_k(\ell)\) is simply the finite-window normalization of the
log-shell weight.  Explicitly,
\(Z_k=(e^{\zeta_k\ell}-1)/\zeta_k\) for \(\zeta_k\ne0\), while
\(Z_k=\ell\) in the marginality.  If \(y=x/\ell\), the same normalized law
depends on \(\zeta_k\) and \(\ell\) only through \(X_k=\zeta_k\ell\):
\begin{equation}
p(y\mid X_k)=\frac{X_k e^{X_k y}}{e^{X_k}-1},
\qquad
p(y\mid0)=1 .
\label{eq:powerlaw_X_collapse}
\end{equation}
Equivalently, the dimensionless normalization factor is
\(Z(X)=\int_0^1 e^{Xy}\mathrm dy\), and the moments of the normalized scale
follow from \(\langle y\rangle=\partial_X\log Z\) and
\(\operatorname{Var}(y)=\partial_X^2\log Z\).  This is why the early
coordinate \(X_k\) is not merely a sign label: it is the full finite-window
coordinate for the normalized scale ensemble.

\begin{figure}[H]
\centering
\includegraphics[width=\textwidth]{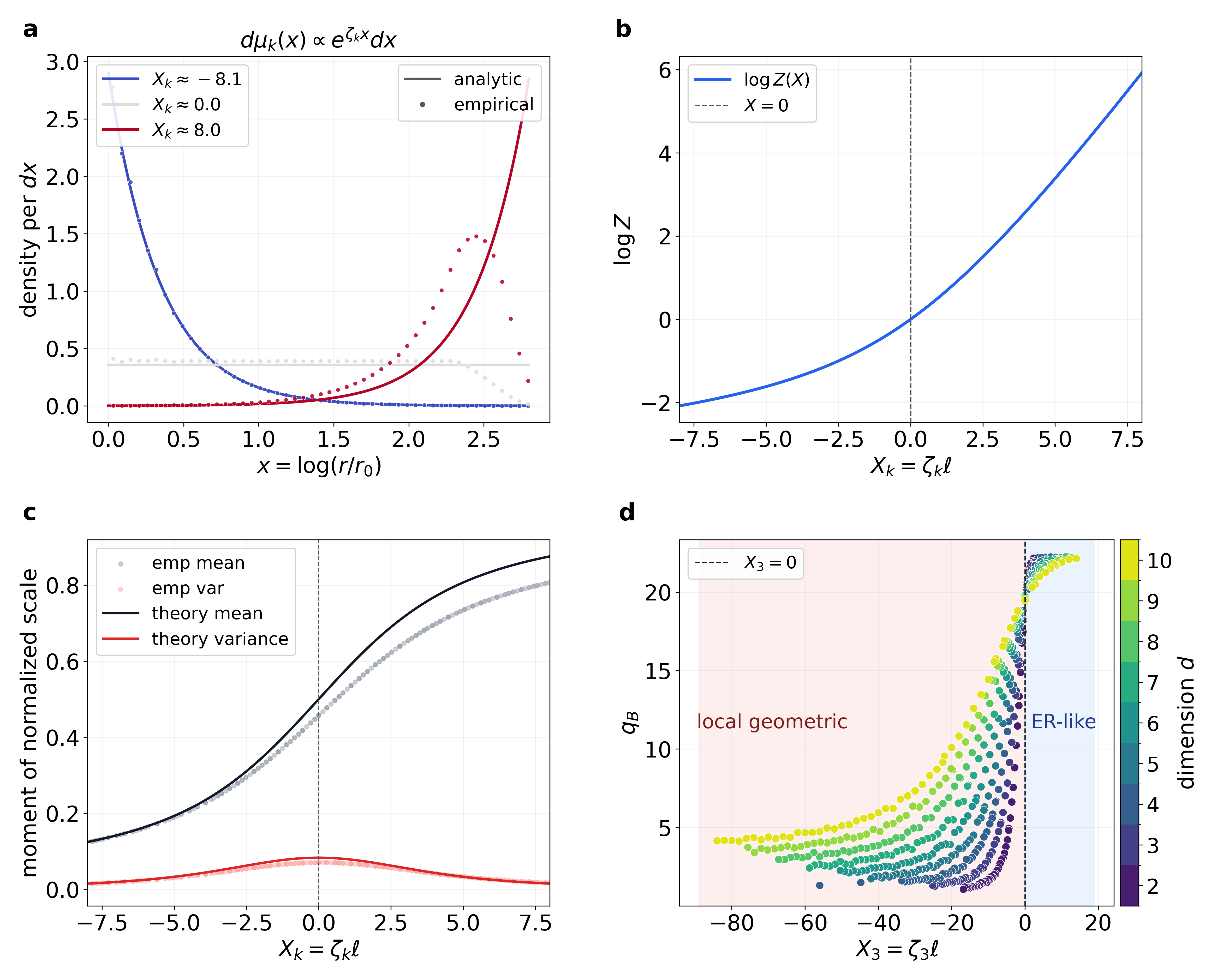}
\caption{Finite-window scale ensemble and its link to local onset.  Panel a
shows the normalized density \(p_k(x)\propto e^{\zeta_k x}\) on the
finite window \(0\le x\le\ell\), comparing analytic curves and empirical
markers for representative negative, zero, and positive values of
\(X_k=\zeta_k\ell\) (\(N=2000,\ d=4,\ n_{\rm stat}=50\)).  Negative \(X_k\) concentrates mass near the core,
\(X_k=0\) gives the uniform log-shell ensemble, and positive \(X_k\)
weights the outer radial shells.  Panel b shows the finite-window
normalizing factor through the analytic function \(\log Z(X)\), with \(X=0\) marking marginality.
Panel c shows the mean and variance of the normalized scale \(y=x/\ell\),
comparing empirical points with the moment formulas derived from \(\log Z\)
(\(N=2000,\ d=4,\ n_{\rm stat}=50\)).
Panel d connects the finite-window coordinate back to percolation by plotting
the local-onset coordinate \(q_B\) against
\(X_3=\zeta_3\ell=d(1-2\eta)\ell\), with color indicating the dimension
\(d\) (\(N=1000,\ k=3,\ 2\le d\le10,\ n_{\rm stat}=50\)).  The vertical line \(X_3=0\) is the finite-window form of 3-clique
closure marginality.}
\label{fig:powlaw-scale-ensemble}
\end{figure}

Figure~\ref{fig:powlaw-scale-ensemble} makes explicit what was used in the
scale-coordinate construction. Panels a--c show that the normalized bare scale law
collapses in \(X_k\): the same coordinate controls the profile shape, the
normalization, and the scale moments.  Panel d is the direct link to the
local-onset threshold for \(k=3\): \(q_B\) is organized by the closure
coordinate \(X_3\), and the reference line \(X_3=0\) is the finite-window form
of 3-clique closure marginality.  This panel is not a separate empirical
summary; it closes the analytic loop. The normalization factor \(Z(X)\) and
its moments define the normalized finite-scale ensemble; \(X_3\) is the same
coordinate specialized for 3-clique closure; and \(q_B\) is the percolation cost
at which 3-cliques become locally reproductive.  The earlier
\((X_2,X_3)\) peak of \(q_\chi-q_B\) and this \(q_B\)-versus-\(X_3\)
relation are
therefore two views of the same mechanism: percolation thresholds are placed on
a finite radial scale ensemble whose natural coordinate is \(X_k\).

\begin{figure}[H]
\centering
\includegraphics[width=\textwidth]{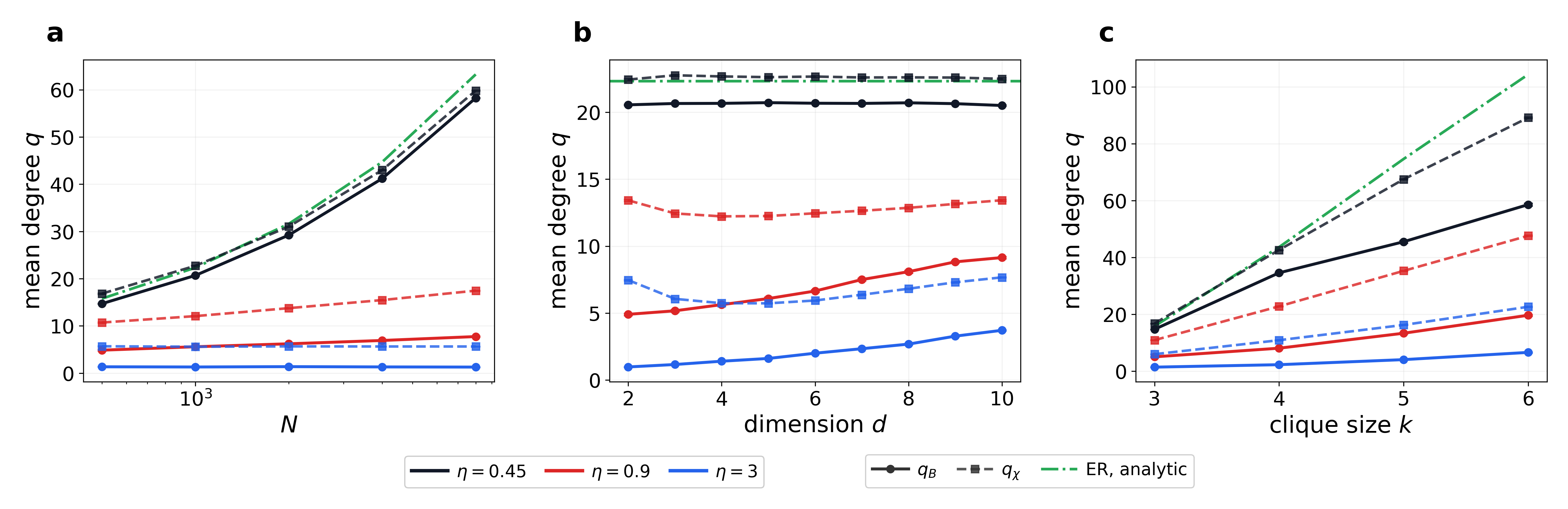}
\caption{Scaling of the power-law percolation picture with \(N\), \(d\), and
\(k\).  Panel a varies system size \(N\), showing
threshold coordinates \(q_B\) and \(q_\chi\) for the representative
\(\eta=0.45,0.90,3.00\) triplet (\(d=4,\ k=3\); \(n_{\rm stat}=50\),
with \(n_{\rm stat}=75\) or \(100\) at the top-up points).  Panel b varies
the dimension \(d\), which
changes the geometric shell-growth rate and therefore the scale coordinates.
The three scale regimes are sampled at fixed system size and clique size
(\(N=1000,\ k=3\); \(n_{\rm stat}=100\) for \(\eta=0.45,0.90\) and
\(n_{\rm stat}=200\) for \(\eta=3.00\)).  Panel c varies the clique size
\(k\), plotting the local
onset \(q_B\) together with the diagnostic finite-response scale \(q_\chi\)
(\(N=500,\ d=4\); \(n_{\rm stat}=180\) for \(\eta=0.45\) and
\(n_{\rm stat}=240\) for \(\eta=0.90,3.00\)).  Error bars
show the available replica uncertainty, and the green reference curves mark
the corresponding Erdos--Renyi analytic thresholds where included.}
\label{fig:powlaw-scaling}
\end{figure}

Figure~\ref{fig:powlaw-scaling} examines the stability of the picture.  Panel
a varies \(N\), changing the accessible scale window and therefore the
finite-window coordinates.  Panel b varies \(d\), changing the geometric
 growth of the shell.  Panel c varies \(k\), which changes the number \(k-1\) of
new kernel factors in a continuation and shifts
the relevant clique-size limit from \(\eta=1/2\) for \(k=3\) to
\(\eta=1/(k-1)\) in general.  The figure is therefore a stability view of the
same mechanism rather than a new threshold definition: thresholds are compared
under changes of system size, dimension, and clique size while the clique-size
scale hierarchy remains the organizing object.

\subsection{Power-law kernel summary}
\label{sec:powerlaw_summary}

The power-law kernel generates a hierarchy of radial scale ensembles.  This
places the power-law clique problem close to the classical long-range versus
short-range crossover problem, but with one essential modification.  In a
standard long-range model, the decay exponent decides whether the
long-range tail is relevant or whether the system crosses over to short-range
behavior. Here, the same decay exponent is filtered by clique-continuation
size: the
relevance variable is \(d-(k-1)\alpha\).  The long-range/local boundary is
, therefore, not a single boundary of the graph; it is a clique-continuation
ladder.  In
the 3-clique example, the graph can be long-range for edges while already
local for clique closures at the same value of \(\eta\).  More generally, the
transition from ER-like randomness to structured geometric locality does not
occur simultaneously for all observables.  It begins with higher-order
clique-continuation events and only later reaches the edges.

Finite systems are organized by the scale coordinate
\(X_k=\zeta_k\ell\), and the finite-window normalization \(Z(X)\) and its
moments show that this is the collapse coordinate of the normalized scale
ensemble.  This also clarifies the Griffiths-like interpretation.  The interval
\(q_B<q<q_\chi\) should not be read only as a phase between two critical
landmarks, it is also the dressed percolation response to a mismatch between
radial scale sectors.  In this sense, the rare-region analog is not imposed by
an external disorder field or by a prescribed modular architecture; it is
produced by the power-law scale hierarchy itself.  Local active clique regions
appear because closure events have already become geometrically localized while
the edge background remains broad.  The same components that enter \(B(q)\) and
\(\chi_{\rm finite}(q)\) can be decomposed by their component-level
contribution, their internal continuation structure, and their finite-sector
response.  These objects are the natural analogs of rare locally active
regions in Griffiths physics.

What is established by the radial theory is the clique-size scale hierarchy.
What is observed in the percolation observables is that this hierarchy
stretches the finite-sector response most strongly when different clique sizes
occupy different scale sectors.  This structure is reminiscent of a
scale ordering across event sizes.  In this precise sense, the power-law kernel is
special because it makes the clique size itself a scale variable.

\section{Exponential kernel}
\label{sec:exponential_kernel}

The complementary smooth, exponentially suppressed family is the soft
exponential kernel,
\begin{equation}
W_\beta(r)=\exp\!\left[-(r/\xi)^\beta\right],
\label{eq:exp_kernel_final}
\end{equation}
However, in the problems studied here its
properties are closer to the step kernel: the central point is the absence of
an algebraic tail, which separates it sharply from the power-law kernel.

The distinction from the power-law case is already visible at the level of
radial event weights.  On scales small compared with the torus size,
\begin{equation}
J_k^{\exp}
\propto
\int_0^\infty r^{d-1}\exp[-(k-1)(r/\xi)^\beta]\,\mathrm dr
=
{\xi^d\over \beta}\,
\Gamma\!\left({d\over\beta}\right)(k-1)^{-d/\beta}.
\label{eq:exp_event_weight}
\end{equation}
The change of variables and the associated event-radius estimate are explicitly written
 in Appendix~\ref{app:exponential_radial_integrals}.
Thus, changing from edges to a \(k\)-clique continuation
changes the typical event radius only by
the finite factor \((k-1)^{-1/\beta}\).  There is no \(N\)-dependent separation
between edges and larger clique continuations and therefore no
analog of the power-law scale coordinate that made clique size itself a
large-scale variable.  Large \(\beta\) approaches the hard step. The smaller
\(\beta\) broadens the kernel, but it broadens all clique sizes within a
short-range class until the tail becomes wide enough so that finite-volume
effects dominate the measurement.

\begin{figure}[H]
\centering
\includegraphics[width=\textwidth]{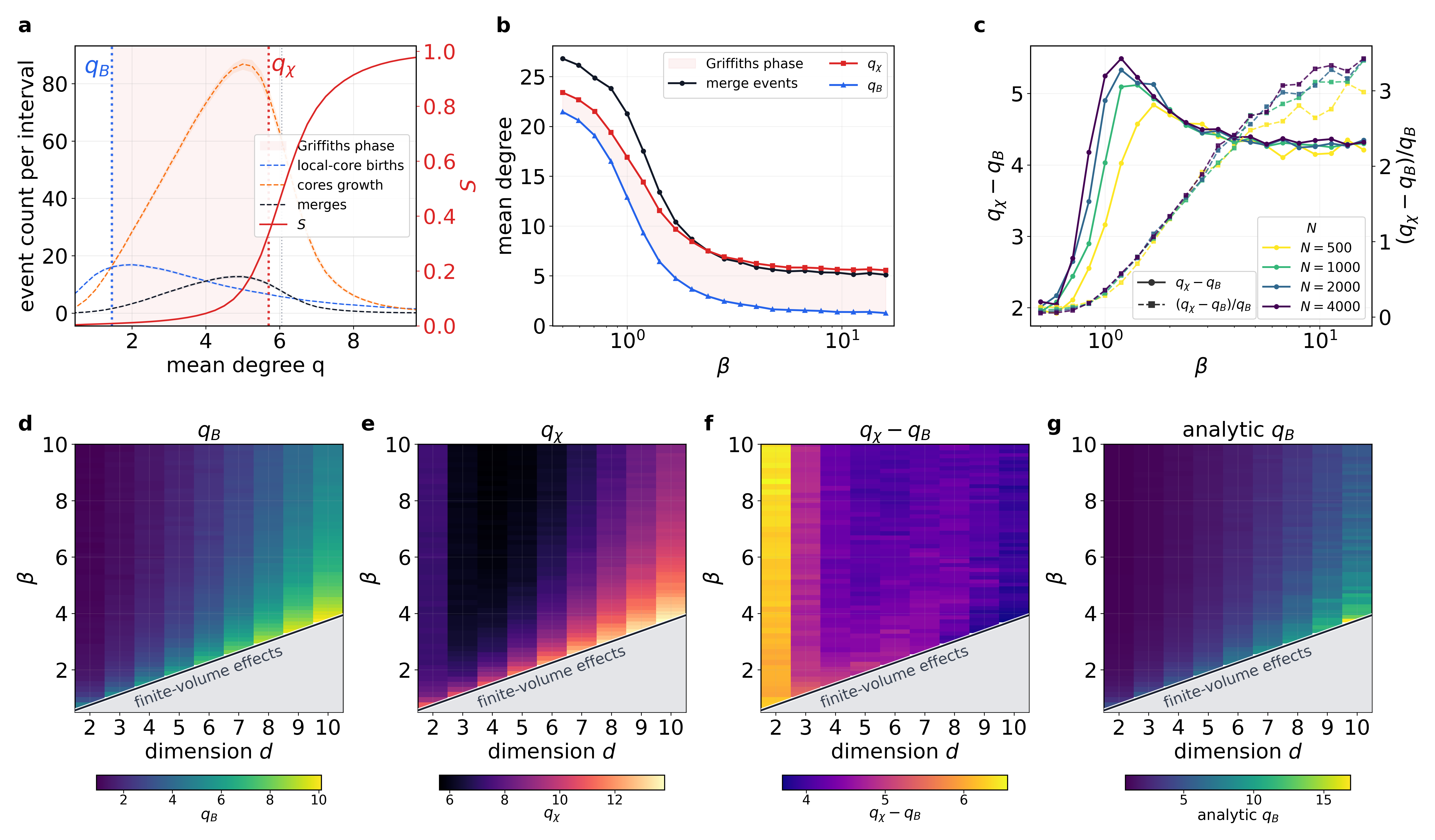}
\caption{Exponential-kernel onset, response width, and \((d,\beta)\) map.
Panel a follows the process-level response through the
waterfall: local \(k\)-clique cores are born at the branching scale, then grow
and merge, while the shaded interval marks the Griffiths-like window between
the local onset and the finite-sector response
(\(N=1000,\ d=4,\ k=3,\ \beta=8,\ n_{\rm stat}=200\)).  Panel b shows the
ensemble-level dependence on the exponential shape parameter \(\beta\) for a
fixed system size, comparing the local branching guide \(q_B\), the
finite-sector guide \(q_\chi\), and the merge-event scale
(\(N=1000,\ d=4,\ k=3,\ n_{\rm stat}=100\)).  Panel c gives the
same response width for several \(N\), both as the absolute separation
\(q_\chi-q_B\) and as the relative separation
\((q_\chi-q_B)/q_B\) (\(d=4,\ k=3,\ n_{\rm stat}=100\)).  The curves show
that the window is stable as a short-range effect rather than amplified into a
new scale hierarchy.  Panels d--g scan dimension against \(\beta\): the first
two maps show \(q_B\) and \(q_\chi\), the third shows the width
\(q_\chi-q_B\), and
the fourth gives the analytic branching prediction for \(q_B\)
(\(N=1000,\ k=3,\ n_{\rm stat}=100\) for panels d--f; panel g is analytic).
The gray
sector below the guide \(\beta_{\rm FV}(d)=0.375d\) is the finite-volume
region, where the exponential tail is too broad relative to the available
periodic box and the measured thresholds should not be interpreted as
infinite-volume behavior.}
\label{fig:exp-process-width-dbeta}
\end{figure}

Figure~\ref{fig:exp-process-width-dbeta} shows the resulting phase diagram in
the same language used for the hard and power-law kernels.  The local
onset \(q_B\) remains below the finite-sector scale \(q_\chi\), so the
Griffiths-like response window survives smoothing of the cutoff.  What is
missing is the extra power-law stretching: the interval is a short-range
component response, not a hierarchy of radial sectors.

\begin{figure}[H]
\centering
\includegraphics[width=\textwidth]{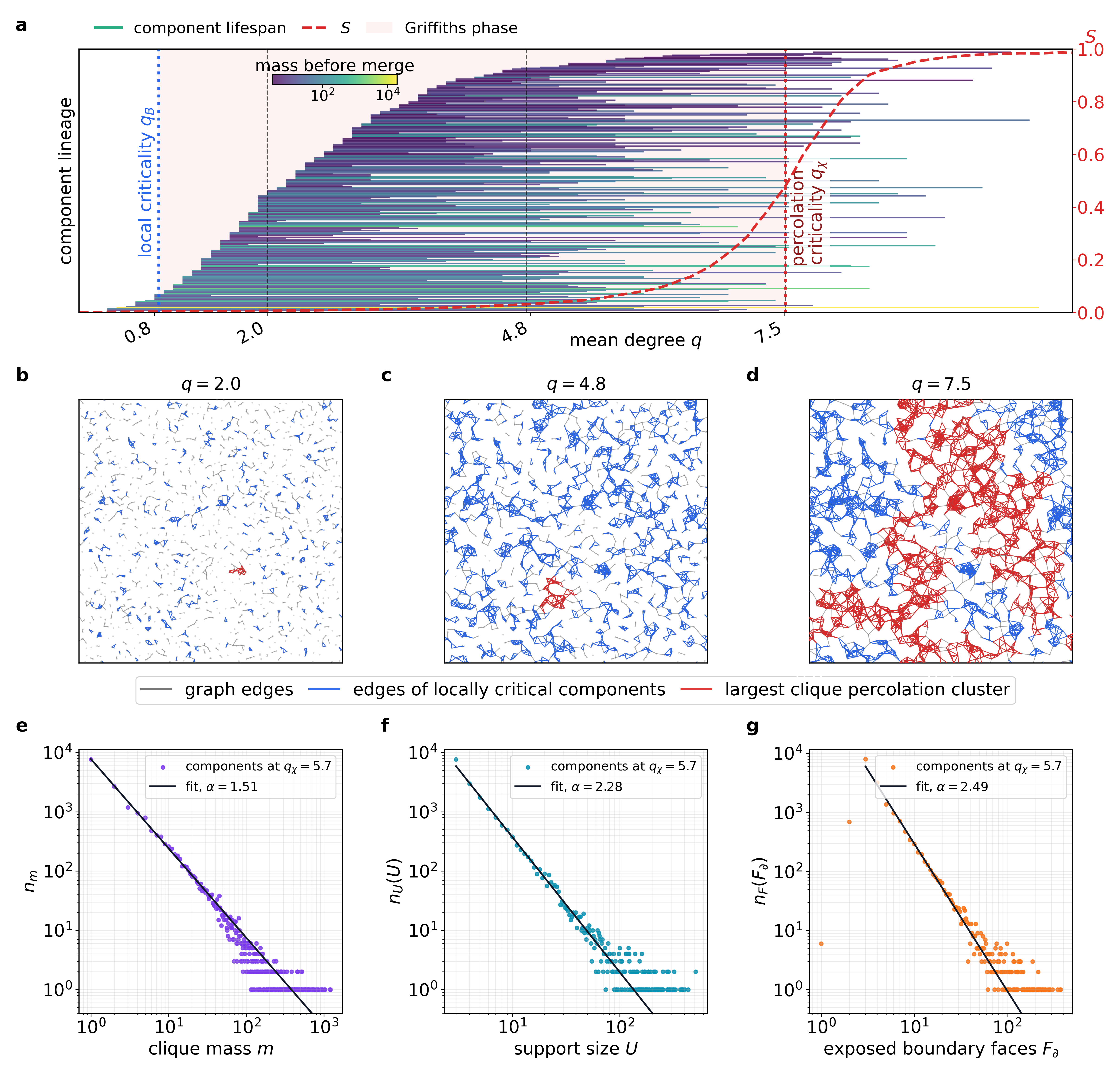}
\caption{Genealogy and finite-component statistics for the exponential
kernel.  The upper part is the exponential analogue of the hard-threshold
genealogy waterfall.  Panel a places the birth, growth, and merge events of
active \(k\)-clique cores on the same \(q\)-axis as the global response, with
the vertical guides marking the local and finite-sector scales.  The graph
snapshots show representative configurations before the window, inside the
window, and close to the global response scale
(\(N=2000,\ d=2,\ k=3,\ \beta=8,\ n_{\rm stat}=100\)).  Periodic geometry is used in
the computation, while the drawing clips boundary-crossing edges so that the
displayed networks retain the local visual geometry.  The lower panels compare
the finite-component distributions at the response scale: component mass,
internal continuation weight, and boundary continuation weight.  The fitted
black curves emphasize that the exponential case has broad finite-component
statistics near response, as in the hard short-range kernel, but these tails
come from critical finite-component organization rather than from an algebraic
interaction tail (\(N=2000,\ d=4,\ k=3,\ \beta=8,\ n_{\rm stat}=100\)).}
\label{fig:exp-genealogy-fincomp}
\end{figure}

The same conclusion appears more directly when one observes clique percolation
and the birth of critical components.  In the hard kernel, the gap between
local activation and global percolation is created by finite clusters that are
already internally active but not yet connected across the system.  The
exponential kernel preserves this picture.  Figure~\ref{fig:exp-genealogy-fincomp}
shows that smoothing the cutoff changes the microscopic ordering of events but
not the qualitative genealogy: active lineages appear after \(q_B\), survive
over a finite interval, and only near \(q_\chi\) become part of the spanning
\(k\)-clique component.

\begin{figure}[H]
\centering
\includegraphics[width=\textwidth]{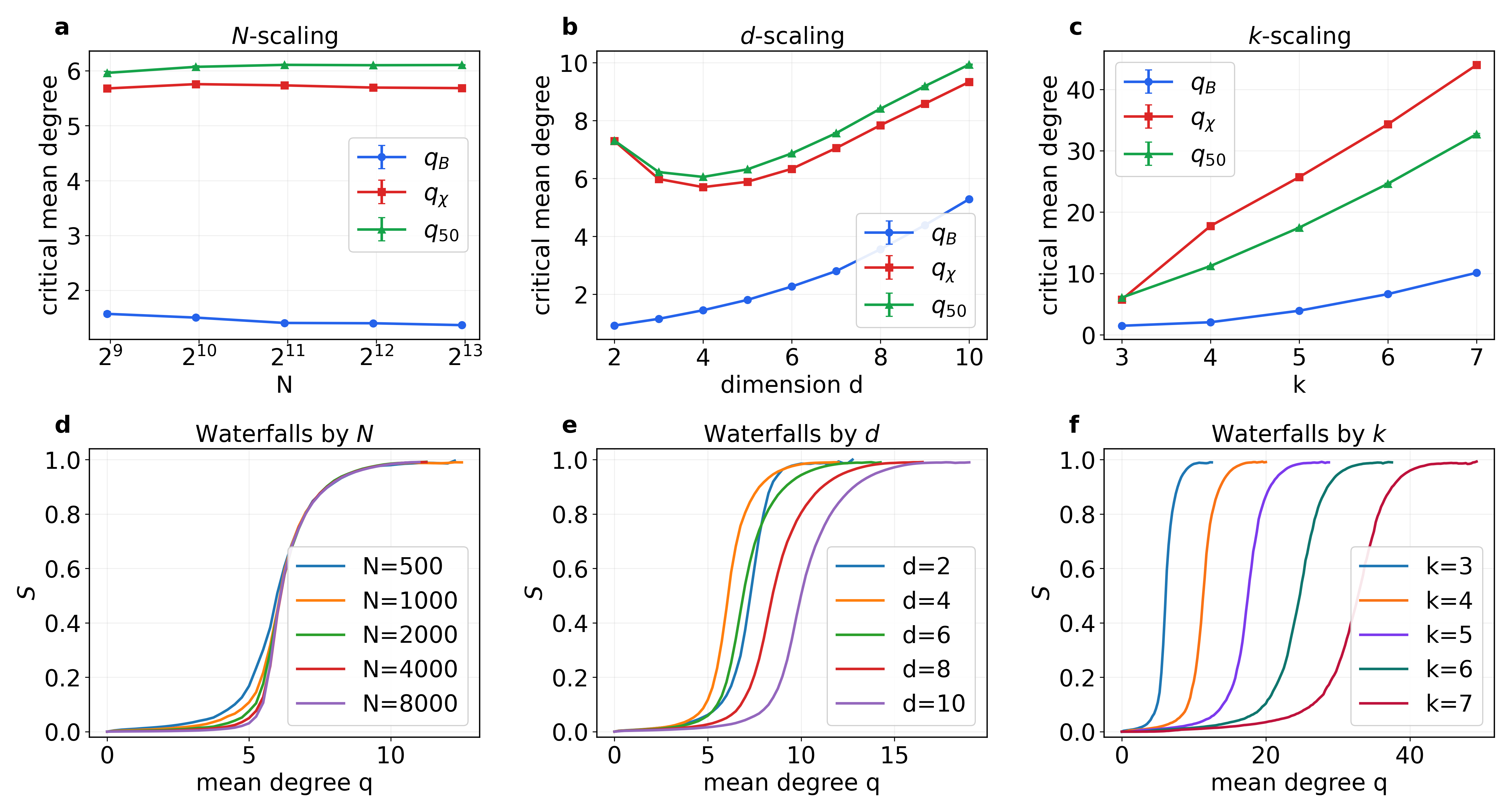}
\caption{Scaling curves and response waterfalls for the exponential kernel.
The top row summarizes threshold trends when one control parameter is varied
at a time: panel a changes \(d\), panel b changes \(k\), and panel c changes
\(N\)
(panel a: \(N=1000,\ \beta=8,\ n_{\rm stat}=200\); panel b:
\(N=1000,\ d=4,\ \beta=8\), with \(n_{\rm stat}=50\) for \(k=3\) and
\(n_{\rm stat}=100\) for \(k=4,\ldots,7\); panel c:
\(d=4,\ k=3,\ \beta=8,\ n_{\rm stat}=100\)).  The
lower row shows the corresponding waterfall curves for the same parameter
directions.  The \(N\)-series confirms that the displayed response window is
not generated by the smallest systems; the \(d\)-series shows a regular
geometric shift of both local and finite-sector scales; and the \(k\)-series
shows the expected displacement of higher-order clique percolation to larger
mean degree.  Together with Figs.~\ref{fig:exp-process-width-dbeta} and
\ref{fig:exp-genealogy-fincomp}, these checks place the exponential kernel in
the same short-range universality class as the hard cutoff, with the
power-law kernel remaining the distinct case where the interaction tail itself
creates the scale hierarchy.}
\label{fig:exp-scaling-waterfalls}
\end{figure}

Finally, Fig.~\ref{fig:exp-scaling-waterfalls} checks that this interpretation
is not a special feature of the single working point.  Varying \(N\) mainly
tests the finite-size stability, varying \(d\) changes the geometric cost of local
closure, and varying \(k\) moves the onset and response to a larger mean degree.
In all three directions, the exponential kernel behaves as a smooth
short-range deformation of the hard model: the waterfalls shift and sharpen
in the expected directions, but no additional regime comparable to the
power-law mixed-scale sector appears.

The exponential kernel therefore completes the comparison.  The split
\(q_B<q_\chi\) is not an artifact of a discontinuous threshold rule; it is a
robust short-range separation between local clique activation and the global
finite-sector response.

\section{Discussion}

In this study, we have investigated the clique percolation in the random 
geometrical graphs with the different geometrical constraints. Our key 
findings are the presence of the Griffiths phase of extended criticality
for all types of constraints and a kind of hierarchy of transitions for 
the power-law kernel. The control parameter for the sharp cut-off and 
exponential kernels is the mean degree of the nodes $q$ and the interval
of values of $q$ supporting the Griffiths phase has been determined 
numerically. The boundaries of the Griffiths phase have a clear-cut
interpretation. One boundary is determined by the local Galton--Watson
condition and is the generalization of the percolation threshold 
for the ER ensemble. Another boundary is determined globally via 
the proper defined susceptibility.

In the Erd\H{o}s--R\'enyi limit, these two notions of criticality collapse
to the same scale: once local clique continuation becomes self-sustaining, a
system-scale clique component emerges.  Geometric correlations separate these
two events.  The local threshold marks the birth of reproducible clique
continuation and the appearance of locally critical clusters, whereas the
finite-sector susceptibility marks the point at which the ensemble of finite
clique components is absorbed by the largest component.  The Griffiths-like
interval is precisely this geometrically induced delay between local activity
and global clique percolation.

For short-range geometric kernels, this delay is controlled by the
structure of finite clique components.  Local active regions appear first,
then grow, expose boundary faces, and connect through the external surfaces of
the clusters before merging into a system-scale component.  This
component-merger mechanism also explains the non-monotonic dependence on
dimension: the critical mean degree is minimized at intermediate dimensions,
around \(d=4\)--\(6\), where finite critical components can merge most
efficiently.

For the power-law kernel the situation is a bit richer although the
Griffiths phase is present as well. Moreover, there is a kind of radial
scale hierarchy for clique continuations of different sizes. The 
effective control parameter in this case depends on the $\eta$ parameter
, and the ordered set of clique-size continuation boundaries replaces the single 
criticality or crossover.

Each clique size has its own separation between ER-like
behavior and the regime where geometric correlations become relevant, and
clique events become local.  The
hierarchy of marginal boundaries is ordered by clique size: larger clique
events become geometrically local earlier than edges.  This produces an
ordered ladder of marginal regimes and scale-free characteristic lengths at
the corresponding crossover scales.  As a result, the graph can still be
nonlocal at the edge level, while clique closures are already geometrically
localized.  This mismatch between scale sectors is what stretches the
Griffiths-like response most strongly in the power-law case.

Our finding provides the ground for the investigation of clique percolation in
real networks which carry one or another geometrical constraint. In
particular, interaction networks are natural examples. For example, the
percolation of monopole clusters has been discussed in
\cite{chernodub2003towards,chernodub2007magnetic} as an indication of the
confinement phase in QCD. In that case, the effective interaction of
instantons--antiinstantons forms a kind of interaction network.

It was recently argued that metric graphs are good approximations for
large-scale human connectomes from the spectral point
of view \cite{bobyleva2025metric} and in percolation analysis
\cite{tiselko2022k}. Therefore, it would be interesting to apply our criteria
in this context and extend the analysis in \cite{tiselko2022k}.
The clique percolation criteria and the boundaries of the Griffiths phase 
can also be useful for the analysis of communication networks.

Finally, note that the geometrical networks are naturally embedded 
in hyperbolic spaces, implying their useful role in the
possible holographic picture. For example, the degree of the node in the
RGG  in d=2 considered as the  boundary of the 3d hyperbolic space 
is defined by the scale of the geometric constraint. On the other hand 
the geometric scale at the boundary in the holographic picture indeed
corresponds to the radial coordinate. On the other hand in hyperbolic 
embedding the node degrees of the network correspond to the radial coordinate as well.

\section{Methods}
\label{sec:numerical_methods}

Each realization was generated on a unit-volume \(d\)-dimensional torus.
The positions of the \(N\) vertex were sampled independently and uniformly from
\([0,1)^d\), and pairwise distances were calculated using the periodic convention. Depending on the vertex positions, the edges were
independent, with connection probabilities determined by the corresponding
radial kernel specified in each kernel section.

For every fixed kernel-shape parameter, we used a monotone coupling in which
each potential edge \((i,j)\) was assigned an independent
\(U_{ij}\sim\operatorname{Uniform}(0,1)\) and activated at scale \(s\)
when \(U_{ij}\leq W_s(r_{ij})\).  Since \(W_s(r)\) does not decrease on the
scanned scale \(R\), \(\xi\), or \(r_0\), this construction produces both the
correct graph distribution on every fixed scale and a nested sequence of
graphs within each realization.  For the hard-threshold kernel, it reduces to
activating edges in increasing order of their torus distance.  Separate
processes were considered for different values of \(\alpha\) and \(\beta\).

We followed the resulting activation order and sampled each trajectory in
prescribed increments of the realized mean degree \(q=2E/N\), where \(E\) is
the current number of edges.  Thus, the sampled values of \(q\) are
fixed-edge-count hitting times of the underlying kernel-scale process rather
than an additional edge probability.  The realized mean degree was used
whenever the numerical critical coordinates were compared with the analytic
mean degree predictions.

Two notions of component size were retained.  The clique mass
\(M_{\mathcal C}\) is the number of \(k\)-cliques in a component
\(\mathcal C\) and is used in the finite-sector susceptibility.  In contrast,
\(S(q)=V(\mathcal C_{\max})/N\) is the fraction of the original graph
vertices belonging to the vertex support of the largest \(k\)-clique
cluster.  Thus, \(S(q)\) measures the macroscopic extent of the largest
component, while \(M_{\mathcal C}\) measures its internal clique content.

For each realization, the local onset \(q_B\) was extracted from the
last upward crossing of the previously defined condition \(B_k(q)=1\), using
linear interpolation between adjacent scan points.  The last crossing was used
to suppress the ambiguities caused by the small finite-sample fluctuations of
\(B_k(q)\).

The numerical estimator corresponding to the theoretical definition of
\(q_\chi\) was obtained by averaging the complete
\(\chi_{\rm finite}(q)\) curves over independent realizations and locating the
maximum of the resulting ensemble curve. The peak positions of individual
realizations were retained separately to quantify finite-sample variability.
Averaging before and after peak extraction are not mathematically identical
operations and were therefore recorded as distinct numerical quantities.
The crossings of fixed levels of \(S(q)\) were used only as descriptive references
and not as definitions of criticality.

To resolve the component dynamics, newly created \(k\)-cliques were classified
by the number of pre-existing \(k\)-clique components that they touched: zero
for a birth event, one for component growth, and two or more for a merger.
When one edge created several \(k\)-cliques simultaneously, they were processed
in a fixed deterministic order. Although the classification of individual events
within a simultaneous activation may depend on the processing order, the total
changes in component masses and susceptibilities are order-independent.

All ensemble averages were calculated from independent realizations with
independently sampled vertex positions and auxiliary edge-uniform variables.
Ensemble curves are reported as means, and error bars or shaded bands represent
standard errors of the mean unless otherwise stated.  Any smoothing of
event-rate curves was used only for visualization; all crossings and peak
coordinates were extracted from unsmoothened data.  Each scan was continued
beyond the response region until \(S\geq0.99\) and \(B_k\geq1\), with one
additional scan step retained after these conditions were satisfied for the first time.

The complete parameter grid, scan spacing, and number of realizations used for each calculation are reported in the corresponding
figure caption.  Finite-volume effects were monitored by comparing the
characteristic kernel scale with the available torus scale.

\appendix

\addcontentsline{toc}{section}{Appendix}
\addtocontents{toc}{\protect\setcounter{tocdepth}{0}}

The appendices follow the order of the article.  We first collect the
kernel-independent finite-component identities, then the hard-threshold
supplementary diagnostics, and finally the power-law and exponential radial
estimates.

\section{General exact component-count and tail-void representation}
\label{app:component_count}

For fixed $N,d,k$, let $n_m(q;N,d,k)$ be the expected number of
$k$-clique components with clique mass $m$.  In a spatial ensemble, this
quantity admits a formal exact representation.  For a soft kernel one must
average not only over the candidate vertex positions but also over the
internal random graph induced on the candidate support.  Let \(H_X\) denote
this internal graph on \(X=(x_1,\ldots,x_u)\).  Then
\begin{equation}
n_m(q;N,d,k)
=
\sum_{u=k}^{N}
{N\choose u}
\int
\mathbb E_{H_X\mid X}
\!\left[
I_{u,m}(H_X,X;q)\,
\bar E(H_X,X;q)^{N-u}
\right]\,
dX .
\label{eq:app_component_support_integral}
\end{equation}
Here \(dX\) is the product position measure and the binomial factor chooses
the candidate support vertices.  The indicator \(I_{u,m}(H_X,X;q)\) enforces
that the internal graph realized \(H_X\) has a \(k\)-clique component whose
vertex support is exactly \(X\) and whose clique mass is exactly \(m\).  This
includes the existence of relevant \(k\)-cliques, connectivity through shared
\((k-1)\)-faces, and the absence of unused support vertices.  The object
\(I_{u,m}\) is an abstract indicator/counting functional that specifies the event
that is being counted; it is not meant to be an explicitly evaluated closed-form
expression.
For a realized internal state \(h\), let \(\mathcal F_\partial(h,X)\) be
the attachable \((k-1)\)-faces of this component.  With the outside-vertex
position integrated against the normalized one-vertex position measure,
\begin{equation}
\bar E(h,X;q)
=
\int
\left[
1-
\mathbb P_z\!\left(
\exists\,Y\in\mathcal F_\partial(h,X):
z\ {\rm connects\ to\ every\ vertex\ of}\ Y
\right)
\right]\,dz
\end{equation}
The probability is that an outside vertex does not attach to the
candidate component. Depending on \(h\) and \(X\), different outside
vertices are independent, which gives the factor
\(\bar E(h,X;q)^{N-u}\).

For the hard-threshold RGG on a unit-volume torus, the same formula
collapses to the deterministic geometric version because \(H_X\) is fixed by
the positions and the radius.  Let $I^{\rm hard}_{u,m}(X;R)$ indicate that the
graph induced by the candidate support $X=(x_1,\ldots,x_u)$ forms a
connected $k$-clique component of the clique mass exactly $m$.  Let
$\mathcal A_X(R)$ be the external attachment region: the set of points from
which an outside vertex would complete at least one attachable
$(k-1)$-face and therefore attach to the component.  Then
\begin{equation}
n_m(q;N,d,k)
=
\sum_{u=k}^{N}
{N\choose u}
\int
I^{\rm hard}_{u,m}(X;R(q))
\left(1-|\mathcal A_X(R(q))|\right)^{N-u}
\,dX .
\label{eq:hard_nm_integral}
\end{equation}
The indicator describes the internal clique-adjacency structure, while the
factor $\left(1-|\mathcal A_X(R(q))|\right)^{N-u}$ is the exact hard-kernel
isolation probability.

The largest-component subtraction requires the probabilities
$\mathbb P(Y_s(q)=0)$.  These are joint tail probabilities, not only averages
of the one-dimensional histogram.  With
\begin{equation}
Y_s(q)=\sum_{m\ge s}N_m(q),
\end{equation}
the event $Y_s(q)=0$ means that the component-count tail above level $s$ is
empty.  Inclusion--exclusion gives
\begin{equation}
\mathbb P(Y_s(q)=0)
=
\sum_{j\ge 0}
\frac{(-1)^j}{j!}\,
\mathbb E[(Y_s(q))_j],
\end{equation}
where $(Y_s)_j=Y_s(Y_s-1)\cdots(Y_s-j+1)$.  The first correction subtracts the
expected number of components with mass at least $s$.  The second adds back
configurations in which two such components coexist and were subtracted twice.
Higher terms continue the same correction for triples, quadruples, and larger
collections.

The factorial moment $\mathbb E[(Y_s)_j]$ counts ordered $j$-tuples of
components with mass at least $s$.  In a geometric graph, each such moment can
be represented by an analytic integral over $j$ mutually isolated component
supports.  This is why the exact largest-component subtraction requires joint
component-count information.  The mean histogram $n_m(q)$ gives the expected
number of large components, but not the probability that several large
components coexist.

\subsection*{Largest-component subtraction and peak balance}

The identity used in the main text follows from the layer representation
\(t^2=\sum_{s=1}^{t}(2s-1)\).  For the largest clique-component mass,
the layer \(2s-1\) is present exactly when \(M_{\max}(q)\ge s\), equivalently
when \(Y_s(q)>0\). Therefore,
\begin{equation}
\mathbb E[M_{\max}(q)^2]
=
\sum_{s\ge1}(2s-1)
\bigl[1-\mathbb P(Y_s(q)=0)\bigr],
\end{equation}
which gives Eq.~\eqref{eq:exact_finite_sector_nm} after subtracting this
largest-component contribution from
\(\mathbb E[\chi_{\rm all}(q)]=\sum_{m\ge1}m^2n_m(q)\).

At an interior differentiable maximum \(q_\chi\),
Eq.~\eqref{eq:exact_finite_sector_nm} gives the exact peak condition
\begin{equation}
\left.
\frac{d}{dq}
\sum_{m\ge 1}m^2 n_m(q)
\right|_{q=q_\chi}
=
\left.
\frac{d}{dq}
\sum_{s\ge 1}(2s-1)
\bigl[1-\mathbb P(Y_s(q)=0)\bigr]
\right|_{q=q_\chi}.
\label{eq:finite_sector_peak_balance}
\end{equation}
Thus, the finite-sector peak is the point where the production of
finite-component second moment is balanced by absorption into the largest
component.

\section{General practical one-large closure}
\label{app:one_large}

The exact finite-sector formula contains the probabilities
$\mathbb P(Y_s(q)=0)$.  These probabilities are determined by the joint law of
large component counts, which is substantially harder to compute than the mean
histogram $n_m(q)$: it requires correlations between large components, not
only their expected numbers.

For this reason, we use a practical closure for the largest-component
subtraction, conditional on a known or measured component-count histogram
$n_m(q)$.  The superscript $\ell$ denotes the one-large closure.  At each mass
level $s$, it treats the occupied tail as contributing at most one effective
largest component.

Define the expected number of components with clique mass at least $s$ by
\begin{equation}
\lambda_s(q)
=
\sum_{m\ge s}n_m(q).
\end{equation}
The one-large closure approximates the tail occupation by
\begin{equation}
\mathbb P(Y_s(q)>0)
\approx
\min(\lambda_s(q),1).
\end{equation}
When $\lambda_s(q)\ll 1$, components of mass at least $s$ are rare, so the
occupation probability is close to the expected count.  Once $\lambda_s(q)$
reaches order one, the tail is already likely to be occupied, and the
probability saturates.  The approximation therefore preserves the rare-tail
behavior while enforcing the probability bound, without introducing an
additional fitted transition level.

The resulting finite-sector approximation is
\begin{equation}
\chi_{\rm finite}^{\ell}(q)
=
\sum_{m\ge 1}m^2n_m(q)
-
\sum_{s\ge 1}(2s-1)\min(\lambda_s(q),1),
\end{equation}
with peak location
\begin{equation}
q_\chi^{\ell}
=
\arg\max_q \chi_{\rm finite}^{\ell}(q).
\end{equation}
The role of $q_\chi^{\ell}$ is limited but useful: it tests how much of the
largest-component subtraction can be reconstructed from $n_m(q)$ alone.  The
exact finite-sector object remains the formula with the true tail-void
probabilities $\mathbb P(Y_s(q)=0)$.

\section{Power-law shell integrals and local-onset scaling}
\label{app:powerlaw_shell_integrals}

This appendix gives the scale estimates used in the main power-law section.
Only the powers of \(N\), \(r_0\), and the kernel exponent are retained; constants
depending only on the torus geometry, the angular measure, or the fixed
dimension \(d\) are absorbed into the scaling equalities below. Thus, the
ordinary equality signs in this appendix denote leading scaling equalities, not
constant-level identities.  Let the outer radial cutoff be
\(L=O(1)\).  For the kernel
\[
W_\alpha(r)=\min\{1,(r_0/r)^\alpha\},
\]
a \(k\)-clique continuation has \(k-1\) kernel factors and radial integral
\begin{equation}
J_k(r_0)
=
\int_0^L r^{d-1}W_\alpha(r)^{k-1}\,\mathrm dr .
\label{eq:app_powerlaw_J_definition}
\end{equation}
The kernel is saturated for \(0<r<r_0\) and has the power-law tail for
\(r_0<r<L\). Therefore,
\begin{equation}
J_k(r_0)
=
\int_0^{r_0}r^{d-1}\,\mathrm dr
+
r_0^{(k-1)\alpha}\int_{r_0}^{L}r^{d-1-(k-1)\alpha}\,\mathrm dr .
\label{eq:app_powerlaw_J_exact_split}
\end{equation}
The core contribution is
\begin{equation}
J_{k,{\rm core}}(r_0)
=
\int_0^{r_0}r^{d-1}\,\mathrm dr
=
\frac{r_0^d}{d}.
\label{eq:app_powerlaw_core}
\end{equation}
For the tail, write \(s_k=d-(k-1)\alpha\).  If \(s_k\ne0\), then
\begin{equation}
J_{k,{\rm tail}}(r_0)
=
r_0^{(k-1)\alpha}
\frac{L^{s_k}-r_0^{s_k}}{s_k}.
\label{eq:app_powerlaw_tail_nonmarginal}
\end{equation}
If \(s_k=0\), the same integral is logarithmic:
\begin{equation}
J_{k,{\rm tail}}(r_0)
=
r_0^d\log(L/r_0).
\label{eq:app_powerlaw_tail_marginal}
\end{equation}
Taking \(r_0\to0\) gives three regimes:
\begin{equation}
J_k(r_0)
=
\begin{cases}
r_0^{(k-1)\alpha}, & (k-1)\alpha<d,\\
r_0^d\log(1/r_0), & (k-1)\alpha=d,\\
r_0^d, & (k-1)\alpha>d.
\end{cases}
\label{eq:app_powerlaw_J_cases}
\end{equation}
The first line is tail-dominated: the outer part of the integral is larger
than the saturated core.  The third line is core-dominated: the tail is
integrable and has the same leading scaling as the core.  The middle line is
the logarithmic marginal case, exactly the condition \(\zeta_k=0\).

The mean-degree coordinate is controlled by the edge integral,
\begin{equation}
q= N J_2(r_0).
\label{eq:app_powerlaw_q_edge}
\end{equation}
For \(k\)-clique continuation, the local branching onset is obtained by
requiring the \(k-1\)-factor continuation to have the expected supply \(O(1)\),
\begin{equation}
N J_k(r_0)=1 .
\label{eq:app_powerlaw_branch_condition}
\end{equation}
We now combine Eq.~\eqref{eq:app_powerlaw_q_edge} with
Eq.~\eqref{eq:app_powerlaw_branch_condition}.  Write
\(\eta=\alpha/d\).

First, suppose \((k-1)\eta<1\), or \((k-1)\alpha<d\).  Then
\(J_k(r_0)= r_0^{(k-1)\alpha}\), so the branching condition gives
\begin{equation}
r_0^{(k-1)\alpha}= N^{-1}.
\label{eq:app_powerlaw_branch_tail}
\end{equation}
Since \(\eta<1\) in this regime, the edge integral is also tail-dominated:
\(J_2(r_0)= r_0^\alpha\). Therefore,
\begin{equation}
q_B
=
N r_0^\alpha
=
N\left(N^{-1}\right)^{1/(k-1)}
=
N^{1-1/(k-1)}.
\label{eq:app_powerlaw_qB_tail}
\end{equation}

At the \(k\)-clique continuation marginal point \((k-1)\eta=1\),
\(J_k(r_0)= r_0^d\log(1/r_0)\).  The branching condition gives
\begin{equation}
r_0^d\log(1/r_0)= N^{-1},
\qquad
r_0^d= \frac{1}{N\log N}
\quad\text{to leading logarithmic order}.
\label{eq:app_powerlaw_branch_marginal}
\end{equation}
The edge integral is still tail-dominated because \(\eta=1/(k-1)<1\) for
\(k>2\). Hence,
\begin{equation}
q_B
=
N r_0^\alpha
=
N\left(r_0^d\right)^{1/(k-1)}
=
N^{1-1/(k-1)}(\log N)^{-1/(k-1)}.
\label{eq:app_powerlaw_qB_marginal}
\end{equation}

Next, suppose \(1/(k-1)<\eta<1\). The integral of \(k\)-clique continuation is already
dominated by the core, \(J_k(r_0)= r_0^d\), while the integral of the edge remains dominated
by the tail, \(J_2(r_0)= r_0^\alpha\).  The branching condition gives
\(r_0^d= N^{-1}\), and therefore
\begin{equation}
q_B
=
N r_0^\alpha
=
N\left(r_0^d\right)^\eta
=
N^{1-\eta}.
\label{eq:app_powerlaw_qB_mixed}
\end{equation}
This is the mixed regime: closure is localizing, but edge supply is still
long-range.

At \(\eta=1\), the edge integral is marginal:
\(J_2(r_0)= r_0^d\log(1/r_0)\), while the closure integral for
\(k>2\) is core-dominated.  With \(r_0^d= N^{-1}\),
\begin{equation}
q_B
=
N r_0^d\log(1/r_0)
=
\log N.
\label{eq:app_powerlaw_qB_edge_marginal}
\end{equation}
Finally, for \(\eta>1\), both edge and closure integrals are core-dominated,
so \(J_2(r_0)= r_0^d\) and \(r_0^d= N^{-1}\). Thus,
\begin{equation}
q_B= N r_0^d= O(1).
\label{eq:app_powerlaw_qB_local}
\end{equation}

Collecting these cases gives the general \(k\)-clique continuation estimate
for \(k\ge3\)
\begin{equation}
q_B
=
\begin{cases}
N^{1-1/(k-1)}, & (k-1)\eta<1,\\
N^{1-1/(k-1)}(\log N)^{-1/(k-1)}, & (k-1)\eta=1,\\
N^{1-\eta}, & 1/(k-1)<\eta<1,\\
\log N, & \eta=1,\\
O(1), & \eta>1 .
\end{cases}
\label{eq:app_powerlaw_qB_general}
\end{equation}
For \(k=3\), closure requires two new kernel factors, and
Eq.~\eqref{eq:app_powerlaw_qB_general} reduces to
Eq.~\eqref{eq:powerlaw_qB_scaling_k3} in the main text.

\section{Exponential radial event integrals}
\label{app:exponential_radial_integrals}

This appendix gives the elementary radial estimate used in the exponential
chapter.  Constants depending only on the angular measure are suppressed.  For
a \(k\)-clique continuation with \(k-1\) kernel factors, the infinite-volume
radial weight is
\begin{equation}
J_k^{\exp}
=
\int_0^\infty r^{d-1}\exp[-(k-1)(r/\xi)^\beta]\,\mathrm dr .
\label{eq:app_exp_J_definition}
\end{equation}
Set \(u=(k-1)(r/\xi)^\beta\).  Then
\(r=\xi[u/(k-1)]^{1/\beta}\) and
\begin{equation}
\mathrm dr
=
{\xi\over\beta}(k-1)^{-1/\beta}u^{1/\beta-1}\,\mathrm du .
\label{eq:app_exp_change_dr}
\end{equation}
Consequently,
\begin{equation}
r^{d-1}\mathrm dr
=
{\xi^d\over\beta}(k-1)^{-d/\beta}u^{d/\beta-1}\,\mathrm du,
\label{eq:app_exp_measure_change}
\end{equation}
and Eq.~\eqref{eq:app_exp_J_definition} becomes
\begin{equation}
J_k^{\exp}
=
{\xi^d\over\beta}(k-1)^{-d/\beta}
\int_0^\infty u^{d/\beta-1}e^{-u}\,\mathrm du
=
{\xi^d\over\beta}\Gamma\!\left({d\over\beta}\right)(k-1)^{-d/\beta}.
\label{eq:app_exp_J_result}
\end{equation}
Thus, changing from edges to a \(k\)-clique continuation changes the radial weight
by a finite factor \((k-1)^{-d/\beta}\).  The associated radius scale changes by
\begin{equation}
r_k \sim \xi(k-1)^{-1/\beta},
\label{eq:app_exp_radius_scale}
\end{equation}
because the exponential cutoff is reached when
\((k-1)(r/\xi)^\beta=O(1)\).  More generally, the normalized moments satisfy
\begin{equation}
{\int_0^\infty r^{d-1+m}e^{-(k-1)(r/\xi)^\beta}\,\mathrm dr
 \over
 \int_0^\infty r^{d-1}e^{-(k-1)(r/\xi)^\beta}\,\mathrm dr}
=
\xi^m(k-1)^{-m/\beta}
{\Gamma((d+m)/\beta)\over\Gamma(d/\beta)} .
\label{eq:app_exp_moment_ratio}
\end{equation}
This is the mathematical reason why the exponential kernel does not create the
power-law clique-size scale hierarchy.  Different clique-continuation sizes
are shifted by finite \((k-1)\)-dependent factors inside one finite radial class,
rather than being separated by an \(N\)-dependent logarithmic scale coordinate.
The infinite-volume estimate is reliable only while the resulting radius scale
is well within the torus window; when \(r_k\) becomes comparable with the
periodic-box scale, the finite-volume corrections discussed in the exponential
figures become the dominant limitation.

\addtocontents{toc}{\protect\setcounter{tocdepth}{2}}

\bibliographystyle{unsrt}
\bibliography{references2}

\end{document}